\documentclass[fleqn,10pt]{wlscirep}
\usepackage[utf8]{inputenc}
\usepackage[T1]{fontenc}

\newcommand{\mcol}{\multicolumn}

\newcommand{\beginsupplement}{%
        \setcounter{table}{0}
        \renewcommand{\thetable}{S\arabic{table}}%
        \setcounter{figure}{0}
        \renewcommand{\thefigure}{S\arabic{figure}}%
     }

\title{Network topology mapping of Chemical Compounds Space}

\author[1,2,*]{Georgios Tsekenis}
\author[3]{Giulio Cimini}
\author[4]{Marinos Kalafatis}
\author[2,5]{Achille Giacometti}
\author[6]{Tommaso Gili}
\author[2,1,5,7]{Guido Caldarelli}

\affil[1]{Institute for Complex Systems, National Research Council, Rome, Italy}
\affil[2]{Department of Molecular Sciences and Nanosystems (DMSN), “Ca’ Foscari” University of Venice, Venice, Italy}
\affil[3]{Physics Department and INFN, University of Rome Tor Vergata, Rome, Italy}
\affil[4]{Department of Microbiology, University of Illinois at Urbana-Champaign, Urbana, Illinois, USA}
\affil[5]{European Centre of Living Technologies (ECLT), “Ca’ Foscari” University of Venice, Venice, Italy}
\affil[6]{Networks Unit, IMT School for Advanced Studies Lucca, Lucca, Italy}
\affil[7]{Rara Foundation -  Sustainable Materials and Technologies ETS, Venice, Italy}
\affil[*]{geotsek@gmail.com}

\begin{abstract}
We define bipartite and monopartite relational networks of chemical elements and compounds using two different datasets of inorganic chemical and material compounds, as well as study their topology. We discover that the connectivity between elements and compounds is distributed exponentially for materials, and with a fat tail for chemicals. Compounds networks show similar distribution of degrees, and feature a highly-connected club due to oxygen. Chemical compounds networks appear more modular than material ones, while the communities detected reveal different dominant elements specific to the topology. We successfully reproduce the connectivity of the empirical chemicals and materials networks by using a family of fitness models, where the fitness values are derived from the abundances of the elements in the aggregate compound data. Our results pave the way towards a relational network-based understanding of the inherent complexity of the vast chemical knowledge atlas, and our methodology can be applied to other systems with the ingredient-composite structure.
\end{abstract}

\begin{document}

\flushbottom
\maketitle

\section*{Introduction} 

The space of chemical compounds comprises hundreds of thousands of different combinations of the over one hundred chemical elements. Such an ample volume was produced by employing several experimental and computational techniques developed for the study of Chemistry over the past centuries. Navigating the vast chemical space is a formidable task and has been the topic of previous research (e.g. \cite{Kirkpatrick2004,leach2007introChemoinformatics}). Motivated by the need to harness the burgeoning complexity of the ever-growing chemicals and materials fields, in this manuscript, we present a constitutive relational network study of inorganic chemistry and materials science, relying on the toolbox of complex networks theory \cite{caldarelli2007book, BarabasiNatPhys2012}.  

In the past, chemical reaction networks have been presented for small numbers of reactants \cite{UnsleberReiher2020}, without addressing the overall complexity of the problem. Furthermore, in materials science, recent efforts have concentrated on faster and cheaper targeted engineering of materials, the so-called Materials Genome project \cite{Jain2013, Batra2021}. Such an approach customarily relies on aggregate statistics. However, incorporating meaningful relational networks can significantly improve the inferential power of statistical approaches, such as materials cartography \cite{Isayev2015}. One network approach has been based on the representation of materials phase diagrams \cite{Aykol2019, Hegde2020}. A different approach was to analyze a set of materials as a network,  according to the cross-correlation of the electronic density of states \cite{Nardelli2021}. Unfortunately, these methods produce fully, or almost fully, connected graphs where all substances are related, which is not very different from an aggregate approach.

Here, we construct and study element-compound networks of extensive catalogues/libraries of chemicals and materials. Furthermore, we successfully model these networks with versatile fitness models derived from maximum entropy methodology. That way, we set large bodies of knowledge onto a new frame of reference, providing novel points of view and enabling further future utility.

\section*{Networks from Data}
We construct relational networks from two different datasets that we sample from two separate databases. The first, CRC, is based on inorganic chemical compounds \cite{crcSet}, and the second, AFLOW, is based on inorganic material compounds \cite{aflowicsdSet} (see Methods). Each of the datasets contains $n_C$ compounds and $n_E$ elements; the specific values are shown in Table \ref{table1}.

\begin{table}[h]
\centering
\begin{tabular}{c|c|c|c|c|c}
\mcol{1}{c|}{dataset}& \mcol{1}{c|}{compounds}  & \mcol{1}{c|}{ elements }  & \mcol{1}{c|}{biLnks} & \mcol{1}{c|}{cLinks} & \mcol{1}{c}{eLinks}  \\
\hline \hline
  &  $n_C$ &  $n_E$ & $L$    & $L_C$ & $L_E$ \\
\hline
CRC &  $3149$ &  $89$& $9118$ & $2117169$ & $1118$    \\
\hline
AFLOW &  $32243$ &  $85$ & $97489$ & $106344446$  & $2956$ \\
\hline \hline
\end{tabular}
\caption{Information about the datasets used to build all the networks. $n_C$ and $n_E$ are the number of compounds and elements. $L$ is the number of bipartite links, while $L_C$ and $L_E$ are the number of monopartite links for compounds and elements, respectively.
}
\label{table1}
\end{table}

\begin{figure*}[h]
\centering
\includegraphics[scale=0.35]{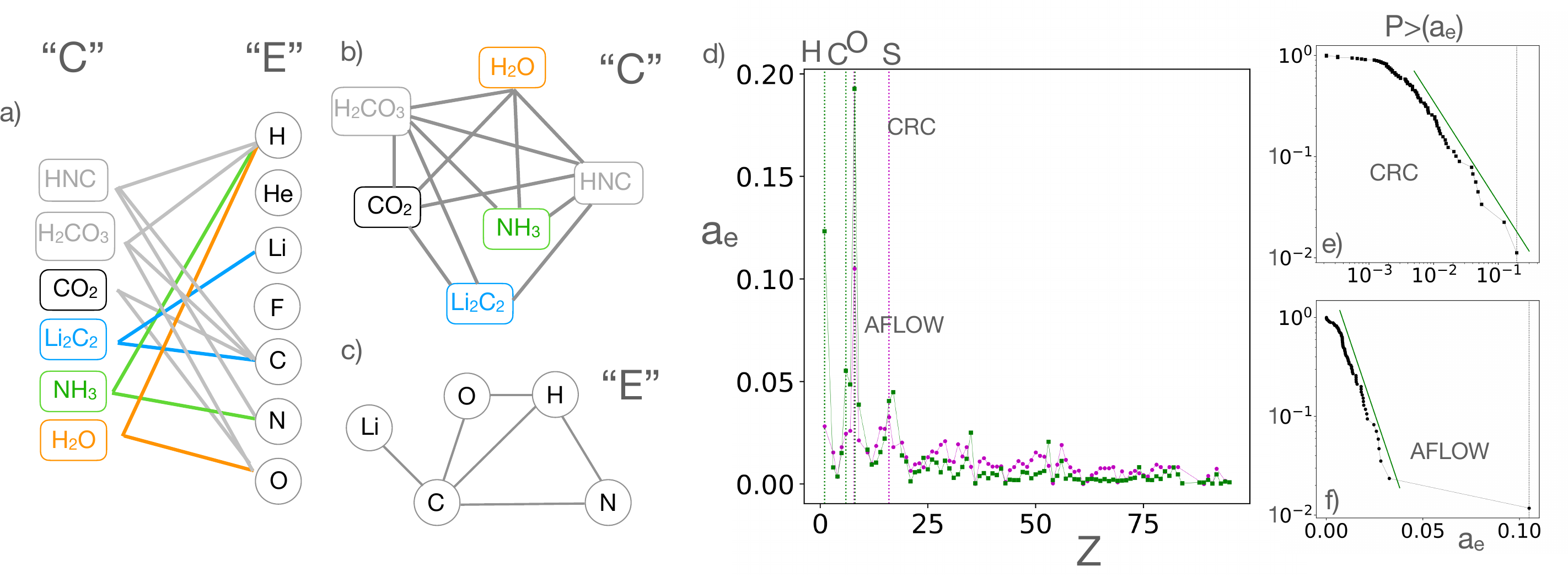}
\caption{Explanatory visualizations of the chemical relational networks, with an illustration of the linking processes on a tiny dataset of six compounds (shown). a) The bipartite network of compounds and elements comprises two layers, layer $C$ for the compounds and layer $E$ for the elements. b) The monopartite projection $C$ of the chemical compounds, which are linked through their common elements. c) Analogously, the monopartite projection $E$ of the elements, which are linked through their co-participation in compounds. d) The relative abundance, $a_e$, vs atomic number, $Z$, of the element species in the CRC (green squares) and AFLOW (purple circles) datasets shows that the most prominent elements are O, H, C for chemicals (green dotted vertical lines), and O, S for materials (purple dotted vertical lines). e,f) The cumulative distribution of element abundances $P_{>}(a_e)$ appears with a fat tail in CRC and an exponential tail in AFLOW. Oxygen is the most abundant element in both datasets.}
\label{fig1}
\end{figure*}

We build a bipartite network for each dataset by linking every compound $c$ to the elements $e$ it contains (Fig. \ref{fig1}a). 
For each dataset, the resulting bipartite network is composed of two layers: one consisting of the compounds $c$ and the other of the elements $e$, and is characterized by a $n_C \times n_E$ binary bi-adjacency matrix $B$ linking the two layers~\cite{Holme2003,Saracco2015,Faust1997,Ings2009,Pavlopoulos2018}, where the matrix element $B_{ce}=1$ if $c$ contains $e$ and zero otherwise
.
For each bipartite network, the total number of links, $L$, is given by the sum of all $B$ matrix elements: $L=\sum_{c,e}{B_{ce}}$. The degree of a node is the sum of its incident links: 
$d_c = \sum_e B_{ce}$ and $d_e = \sum_c B_{ce}$ for compounds and elements, respectively.

\begin{figure*}[h]
\centering
\includegraphics[scale=0.30]{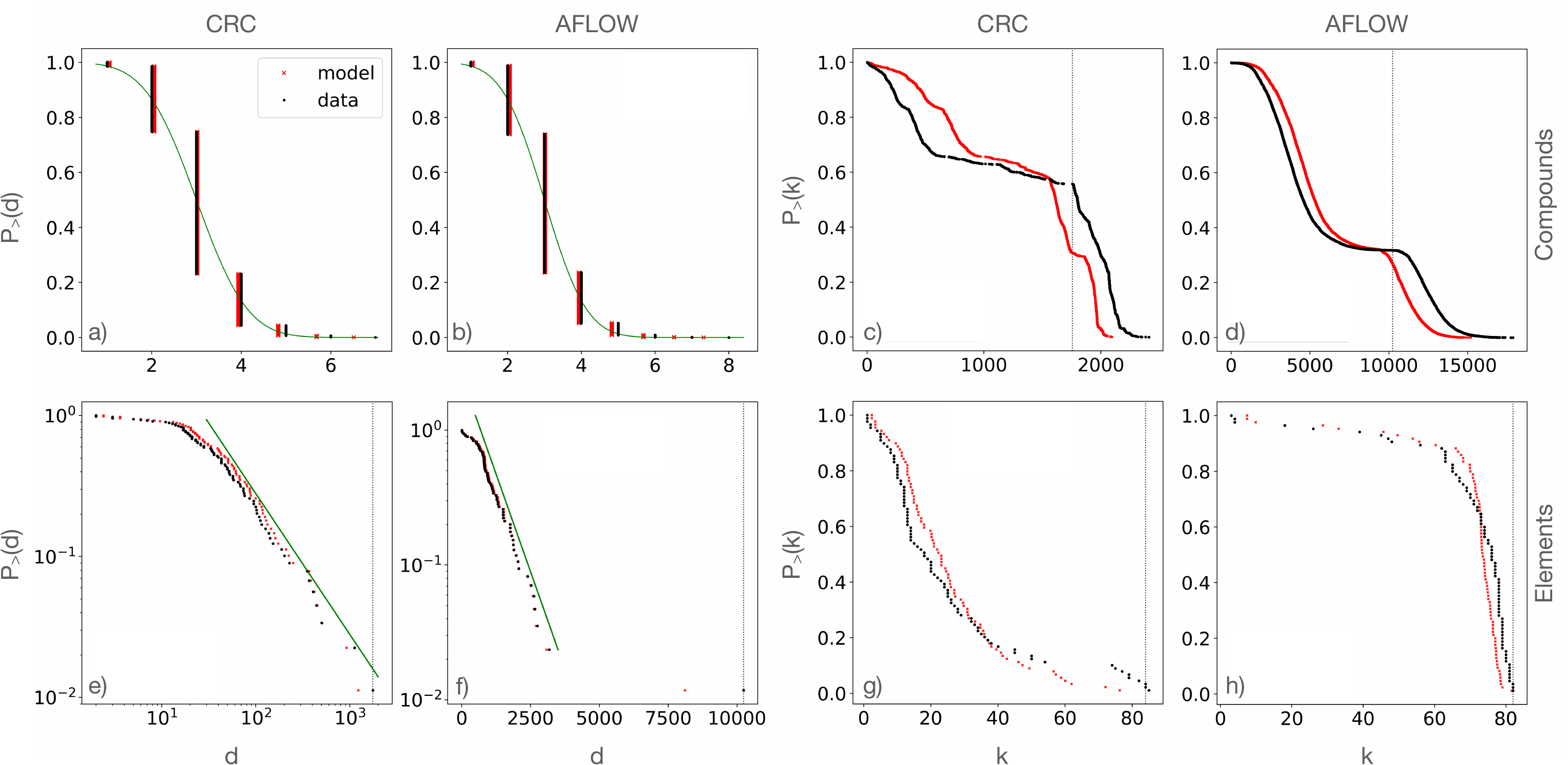}
\caption{Degree distributions. Cumulative degree distribution, $P_{>}(d)$, for the compounds (a,b) and elements (e,f) layers of the CRC and AFLOW bipartite networks. Cumulative degree distribution, $P_{>}(k)$, for the compounds (c,d) and elements (g,h) monopartite networks of CRC and AFLOW. In (c,d) the vertical dotted lines indicate the smallest degree of a compound with oxygen. In (e,f,g,h) the vertical dotted lines indicate the degree of the oxygen element. The continuous green lines in (a,b) are cumulative normal distributions, in (e) a power law $\sim x^{-1.0}$, and in (f) an exponential curve $\sim \exp(-const * x)$. All green lines are visual guides. Black points represent empirical data while red points are obtained from fitness model estimates.
}
\label{fig:Km}
\end{figure*}

The degree distributions for both layers and both datasets are shown in the four left panels of Fig. \ref{fig:Km}(a,b,e,f). The degrees ($d_c$) of the compounds layer are discrete since each compound is linked to as many distinct elements it contains. Their overall distribution appears modulated by a Gaussian-like curve. 
The degree ($d_e$) distributions of the elements appear dominated for larger values by a fat-tail for the CRC network, and by an exponential decay for the AFLOW network, indicating a different inherent complexity of inorganic chemicals vs materials. Oxygen is the most connected element corresponding to the maximum degree, a feature confirmed also by other analyses as we shall see below. 

We further consider the relationships between compounds or between elements, by projecting the bipartite network on either layer to get the corresponding monopartite network. In the compounds network (Fig. \ref{fig1}b), the nodes are the compounds, and a pair of compounds are linked if they share a common element. In the elements network (Fig. \ref{fig1}c), the nodes are the elements, with links between elements that co-participate in a compound.
The adjacency matrices of the binary monopartite networks, $A_C$ and $A_E$, are obtained by the binary bi-adjacency matrix $B$: 
$(A_C)_{cc'}=1$ if $\sum_e B_{ce}B_{c'e}>0$, $(A_E)_{ee'}=1$ if $\sum_c B_{ce}B_{ce'}>0$, and zero otherwise.
Summing all non-zero entries in the adjacency matrices gives the number of links in the monopartite networks: $2L_C = \sum_{c,c'} (A_C)_{cc'}$ and $2L_E = \sum_{e,e'} (A_E)_{ee'}$. The degree of compounds and elements of the monopartite networks are respectively given by $k_c = \sum_{c'} (A_C)_{cc'}$ and $k_e = \sum_{e'} (A_E)_{ee'}$.

The degree distributions of the monopartite compounds ($k_c$) and elements ($k_e$) networks are shown in the four right panels of Fig.~\ref{fig:Km}(c,d,g,h) for both CRC and AFLOW. A striking feature of the degree distributions for compounds is that they appear to be composed of two main modes. Further investigation reveals that all the compounds in the high-degree bump contain the oxygen element. We denote with a vertical black dotted line (upper panels) the smallest degree of a compound that contains oxygen. Correspondingly, in the elements networks oxygen has the maximum, or nearly maximum, degree, which we denote with a vertical black dotted line (lower panels). In both datasets 
we discover that compounds containing oxygen form an oxygen club. This is a maximally interconnected community composed of compounds with a large degree. The oxygen club is a result of oxygen's prominence in the inorganic chemistry and materials science datasets, as well as the rules of the network. This particular feature is almost impossible to be captured by a specific model and needs to be addressed by further analysis of the structure of the datasets. The shape of the degree distributions of the elements networks appears to have an exponential body for the CRC network and a linear decay for the AFLOW network.

\section*{Fitness Models}
We model these networks by assuming that there is a hidden underlying process where all the elements compete for prominence based on an unknown intrinsic fitness. We discover that the abundance, $a_e$, of each element $e$, which is simply the element occurrence in all compounds of a dataset shown in Fig. \ref{fig1}d, is an excellent quantity to consider as element fitness. Similarly, we find that the fitness of a compound $c$ can be represented well by the number of element species, $\ell_c$, it contains.
We model the bipartite networks by using normalized fitness values
\begin{eqnarray}
x^b_{e} = \frac{ a_{e} } { \sum_{e'} a_{e'} }, \quad
y^b_{c} = \frac{ \ell_{c} } { \sum_{c'} \ell_{c'} }
\end{eqnarray}
as effective parameters of a maximum-entropy fitness model~\cite{caldarelli2002scale-free,Garlaschelli2004,Cimini2015,Saracco2015,Squartini2017,Cimini2019,Cimini2022}. 

Specifically, in our model, each pair of nodes from the two different layers (i.e., an element $e$ and a compound $c$) is connected according to a linking probability with a Fermionic form
\begin{eqnarray}
f(\delta,x^b_e,y^b_c) = \frac{ \delta x^b_e y^b_c }{1 + \delta x^b_e y^b_c}
\label{eq:flf}
\end{eqnarray}
where $\delta$ is a single tuning parameter for each network. The best fitting value $\delta^{*}$ is extracted by matching the number of links, $L$, of the real network with that of the model:
\begin{eqnarray}
L = \sum_{e,c} f(\delta^{*},x^b_e,y^b_c) 
\label{eq:Lbi}
\end{eqnarray} 
Using $\delta^{*}$ from Eq.~\eqref{eq:Lbi} and the normalized fitness $\{x^b_e\}$, $\{y^b_c\}$, we calculate the expected model degrees, $\Tilde d_c = \sum_{e} f(\delta^*,x^b_e,y^b_c)$ and $\Tilde d_e = \sum_{c} f(\delta^*,x^b_e,y^b_c)$.

We remark that Eq.~\eqref{eq:flf} derives from an entropy maximisation procedure with degree constraints, where the fitness values replace the unspecified Lagrange multipliers~\cite{Park2003,Park2004,Cimini2019}. 
Hence our modelling is an effective maximum-entropy procedure informed by a heuristic fitness ansatz, where the fitness of the nodes generate the model degrees. 
The alternative route, which we do not follow here, would be to find the values of the multipliers such that the expected degrees match the empirical values, through e.g. likelihood maximization. 

For the monopartite networks, we follow a similar approach. We use abundance for the fitnesses of the elements, while for the compounds we sum up the abundances of the elements they contain, $a_{c} = \sum_{e \in c} a_{e}$. Hence
\begin{eqnarray}
x^m_e=\frac{ a_{e} } { \sum_{e'} a_{e'} }, \quad y^m_{c} = \frac{ a_{c}} { \sum_{c'} a_{c'} }.
\end{eqnarray}
Links between nodes in the two monopartite networks are computed with a linking function similar to the previous one. We have, for elements and compounds networks respectively
\begin{eqnarray}
f(\delta_E,x^m_e,x^m_{e'}) = \frac{ \delta_E x^m_e x^m_{e'} }{1 + \delta_E x^m_e x^m_{e'}}, \qquad
f(\delta_C,y^m_c,y^m_{c'}) = \frac{ \delta_C y^m_c y^m_{c'} }{1 + \delta_C y^m_c y^m_{c'}}.
\end{eqnarray}
$\delta_E$ and $\delta_C$ are still free parameters for each network, whose values $\delta_E^{*}$ and $\delta_C^{*}$ are determined using the number of links in the empirical networks:
\begin{eqnarray}
2L_E = \sum_{e,e'} f(\delta_E^{*},x^{m}_e,x^{m}_{e'}), \qquad
2L_C = \sum_{c,c'} f(\delta_C^{*},y^{m}_c,y^{m}_{c'})
\label{eq:2Lmono}
\end{eqnarray}
Again, we calculate the expected degrees, $\Tilde k_e$, $\Tilde k_c$, using $\delta^{*}_E$, $\delta^{*}_C$ from Eq. \eqref{eq:2Lmono}, and the normalized fitness $\{x^m_e\}$, $\{y^m_c\}$, respectively, using $\Tilde k_e = \sum_{e'} f(\delta_E^{*},x^{m}_e,x^{m}_{e'})$ and $\Tilde k_c = \sum_{c'} f(\delta_C^{*},y^{m}_c,y^{m}_{c'})$.


We find very good or exceptional agreement between the real networks and the fitness models regarding the degrees in all cases, as shown in Fig.~\ref{fig:Km}. The higher-order network measure of the degree assortativity exhibits stronger fluctuations but is still captured on average as we report in the SI and Figs.~\ref{fig:KnnvsK} ~\ref{fig:KnnvsKm}.

\section*{Community analysis}

\begin{figure*}[h]
\centering
\includegraphics[scale=0.35]{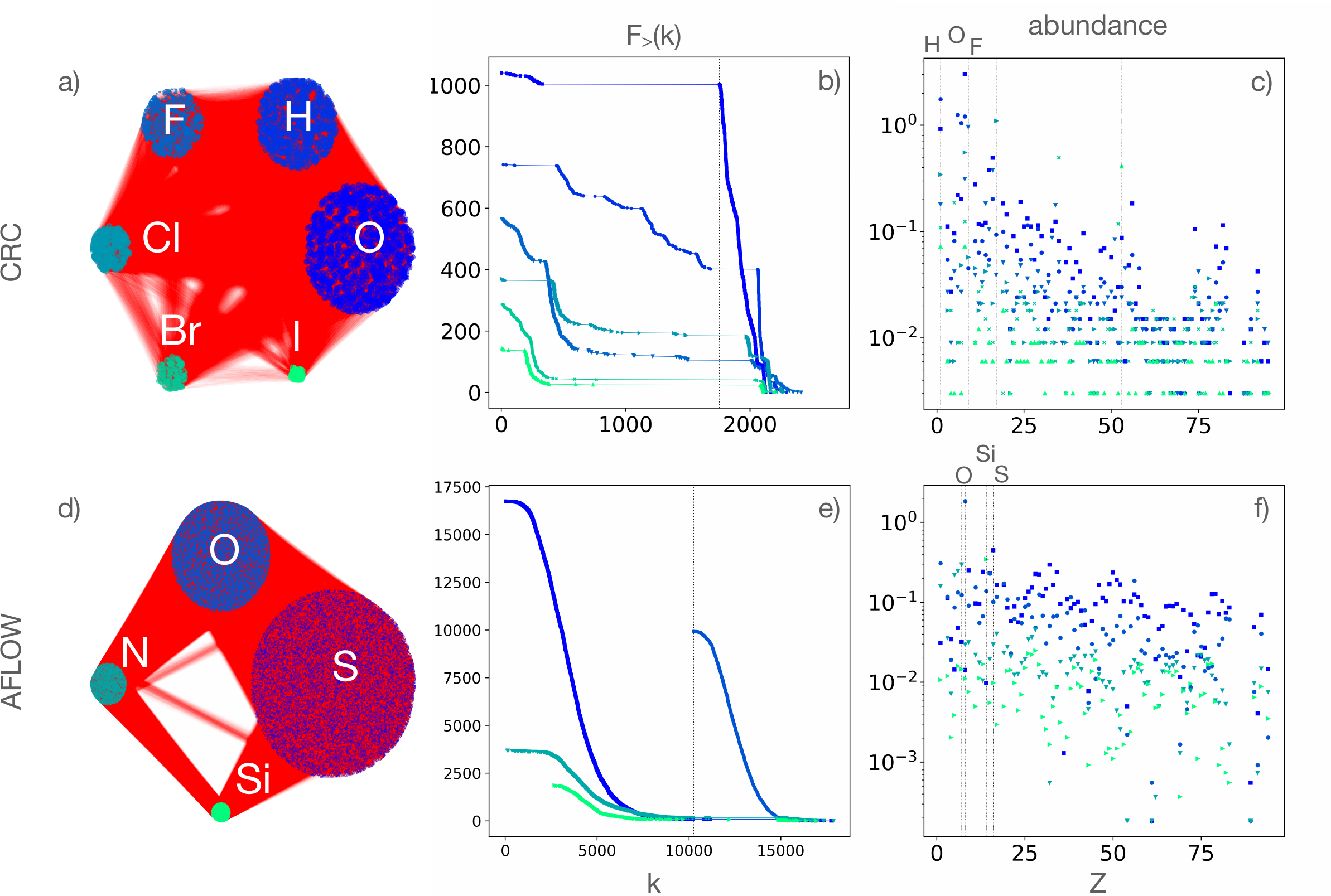}
\caption{(a,d): Community structure of the compounds monopartite networks, for CRC and AFLOW; each community is labeled by its dominant element. Communities are colored according to their size (blue to green representing largest to smallest groups respectively), and the links are red. The most connected compounds ordered by descending community size are, for CRC: $Na_{3}PO_{3}S\cdot12H_{2}O$, $(C_{2}H_{5}O)_{2}P(S)SNH_{4}$, $NH_{4}SO_{3}F$, $CuCl_{2}\cdot2NH_{4}Cl\cdot2H_{2}O$, $NH_{4}BrO_{3}$, $NH_{4}IO_{3}$,
and for AFLOW: $C_{4}F_{12}N_{8}O_{12}S_{8}Se_{8}$ (5936f8e3995c49e8), $Co_{2}H_{48}Ni_{2}O_{40}S_{4}$ (b730426d3f44aa15), $C_{12}Cl_{12}N_{12}O_{4}P_{4}S_{12}Sb_{4}$ (155cbacf430e2898), $Ca_{12}F_{1}K_{1}O_{26}S_{2}Si_{4}$ (3d134e8d4260e63e). In parenthesis we give the AFLOW unique identifiers for each compound. 
(b,e) Cumulative counts of degree, $F_{>}(k)$, for each community; the vertical dotted line indicates the smallest degree of compounds with oxygen. (c,f) Relative abundance of element species in communities, where the most abundant elements are indicated on the top of the plots. 
}
\label{fig:Comms}
\end{figure*}

We further analyze the community structure emerging from this way of exploring the chemical space. We use the Louvain greedy algorithm \cite{Nguyen2008}, a method based on the maximization of the modularity $Q$ (a quantity related to how many links tend to connect nodes within communities rather than nodes belonging to different communities \cite{NewmanGirvanPRE2004, NewmanPNAS2006,  Good2010, Zhang2014, Bongiorno2017}). We identify between 3 and 5 communities in AFLOW, with $Q \approx 0.25$, and between 5 and 7 communities in CRC, with a smaller $Q \approx 0.11$, as shown in Fig. \ref{fig:Comms}. The small variability of the results depends on the initialization of the algorithm; below we discuss only the findings that are robust across multiple runs of the algorithm.

As expected, there is a community of compounds of large degree, which has the highest abundance of oxygen ($Z=8$), (Fig.\ref{fig:Comms} a,d). The oxygen community has a high overlap with the oxygen club, but they are not identical,  (Fig.\ref{fig:Comms} b,e). The rest of the communities are centered around other, not necessarily prominent elements,  (Fig.\ref{fig:Comms} c,f). More specifically, in the CRC network, the second largest community is dominated by hydrogen ($Z=1$), and the third by fluorine ($Z=9$). We notice that the three most prominent elements in the CRC dataset overall, O, H, C ($Z=6$) (Fig.\ref{fig1}d), and the most prominent elements of the three largest communities, are all light elements (first row of their groups in the periodic table) and are highly reactive. In the AFLOW network there are communities that contain most of the oxygen ($Z=8$), sulfur ($Z=16$) and silicon ($Z=14$). We notice that two most prominent elements in the AFLOW dataset, O and S (Fig.\ref{fig1}d), have their own communities, and are the first two elements of the original group $VII$ or the newer group $16$ of the periodic table, which are collectively called chalcogens.

\section*{Discussion}
In summary, we developed a simple but fundamentally effective way to delve into the hidden complexity of large, aggregate, chemical datasets, and reveal their higher-order correlations. 
We discovered that the connectivity of elements to compounds follows a heterogeneous  distribution with different kinds of tails: a fat one for the CRC network and an exponential one for the AFLOW network. We traced this significant difference to the corresponding distributions of elemental abundance in the CRC and AFLOW datasets, as shown in Fig.\ref{fig1}(e,f). The connectivity analysis also revealed the special role of oxygen in the networks as we found that it dominates all orders of correlation amongst inorganic AFLOW and CRC, Fig.\ref{fig1}. Therefore, we revealed that oxygen holds a prominent position in the complexity of inorganic chemistry, beyond simply being the most common element \cite{Wang2021} (Oxygen has also been found to play  a central role in biochemical networks and the complexity of life \cite{RaymondSegreScience2006}). A further community analysis we performed revealed chemical knowledge of purely topological origin. The largest communities in CRC compounds network are dominated by light, highly reactive elements. The picture is starkly different for AFLOW, where the most prevalent elements are somewhat heavier and less reactive. The AFLOW compounds network is less modular, comprising more communities, as compared to the CRC. All of the results presented in this Report are obtained thanks to the network methodology we developed, and cannot be derived from aggregate analyses. 

In addition, we were able to formalize our findings through a maximum entropy network approach. Our fitness models were tailored for the bipartite network and its monopartite projections, employing a single-fitted parameter and novel fitness values that are external to the network. Our analysis is able to quantify self-consistently both networks of CRC and AFLOW, and reproduces successfully their statistically different connectivity. The parsimonious modelling methodology we developed can be applied to any bipartite network, or to a pair of complementary networks, such as article-author networks~\cite{NewmanPNAS2001}, recommendation networks~\cite{Zhou2007}, disease phenome-genome~\cite{ Goh2007}, countries-products~\cite{Hidalgo2009,Tacchella2012,Saracco2015}, food ingredients-flavors~\cite{Ahn2011}, social networks~\cite{Faust1997, Holme2003}, ecological networks~\cite{Ings2009}, biological and medical networks~\cite{lee2008implications, zhou2014human, Pavlopoulos2018}, and so on.

Network science can benefit chemistry and materials science by reorganizing its extensive body of knowledge through complex networks. Analyzing and modeling chemistry networks allows us to systematize intrinsic behaviors and emergent or occluded patterns into quantitative relations. Such informed chemical/material graph atlases can accelerate decisions on "synthesizability", and minimize costs for intelligent design of novel composites with desired properties. This can be done by utilizing graphical algorithms and network methods to complete tasks that are computationally overwhelming or demanding to investigate as is the case when starting from raw data, first principles, experimentally, or traditional cheminformatics~\cite{leach2007introChemoinformatics}. Specifically, the network connectivity properties that we study here describe the relation among existing substances, and can inform searches for alternative or novel ones.

Our approach involves large networks of substances, different from approaches that perform learning with neural networks on individual, modestly-sized, molecular graphs~\cite{NIPS2015_f9be311e,pmlr-v70-gilmer17a}. In its current form, it takes advantage solely of the chemical composition of substances, but it can be systematically expanded to include more material properties as node variables, such as crystal structure, thermodynamic quantities, or mechanical properties. It can utilize more sophisticated measures for weighted linking, e.g. the number of atoms in common, or quantify the similarity of nodes with cosine/dot-product, for further gains.

\section*{Methods}

\textbf{Datasets creation from databases} 

From the AFLOW database \cite{aflowicsdSet} we downloaded all the compounds that also belong to the ICSD catalogue (similarly to \cite{Nardelli2021}), and have a value entry in the following eleven properties: composition, species, density, volume atom, pressure, valence cell iupac, spin atom, scintillation attenuation length, energy atom, enthalpy atom, eentropy atom (electronic entropy). 

We utilized the entire database of Physical Constants of Inorganic Compounds of the CRC Handbook of Chemistry and Physics Online 102nd Edition (2021), which is part of CHEMnetBASE \cite{crcSet}. 

Both databases may reflect the biases of their creators, historical trends in chemistry, and/or the research interests, needs, and abilities of the scientific and engineering community. The AFLOW database comprises solids, while the CRC database contains compounds in all phases at standard conditions complicating property annotation. The only implicit constraint we imposed on the AFLOW database was the materials to be sufficiently well analysed/studied. We presume that the CRC database was compiled with a similar criterion of most commonly used substances. Our results are proven for our datasets, since the global shapes of the distributions are preserved when we randomly sub-sampled our datasets. As the databases expand and research becomes gradually more systematic, we foresee that the potential benefit from our network analysis would also expand past the space limited by current research.\\

\textbf{Graph link density}

The link density of the networks of elements is an increasing function of the size of the compounds datasets considered, while the density of the networks of compounds is independent of such size. This is due to the fact that for the elements network, the total number of nodes (i.e. elements) is constant, $n_E \sim O(100)$, while the number of links between elements increases as more compounds are analyzed. For the compounds network, the length  of compounds. i.e. the number of elements species, is nearly constant, $1 \leq \ell_c \leq 8$, and as more compounds are added, both the number of nodes (i.e. compounds) and links increase. This results in a constant link density, which is roughly $\sim 0.20$ for the AFLOW and $\sim 0.42$ for the CRC compounds networks (see SI, Fig.~\ref{fig:linkDensity}
). 

\section*{Acknowledgements}
We would like to thank Stefano Bonetti and Roberta Sinatra for useful discussions. This work has been partially supported by the grant EU ’HumanE-AI-Net’, no. 952026, and by the Italian Ministry of Foreign Affairs and International Cooperation (“Mac2Mic”), and by MIUR PRIN-COFIN2022 grant 2022JWAF7Y. 

\section*{Author contributions statement} GT and GCa conceived the paper, designed research, performed research, wrote the manuscript. GT conducted the analysis of the empirical data and the numerical modeling. GCi, MK, AG, TG, contributed to research, data, and writing of the paper.

\section*{Additional information}
\textbf{Competing interests} The authors declare no competing interests.

\section*{Data Availability}
The CRC dataset can be obtained from the table of Physical Constants of Inorganic Compounds of the CRC Handbook of Chemistry and Physics Online 102nd Edition (2021), which is part of CHEMnetBASE \cite{crcSet}, at https://hbcp.chemnetbase.com . \\
The AFLOW dataset can be obtained from the AFLOW library of crystallographic prototypes~\cite{aflowicsdSet}, at http://www.aflowlib.org .

\bibliography{six}

\newpage
\clearpage

\beginsupplement

\section{Supplementary Information}

\subsection{Link density of networks}

\begin{figure*}[th]
\centering
\includegraphics[scale=0.5]{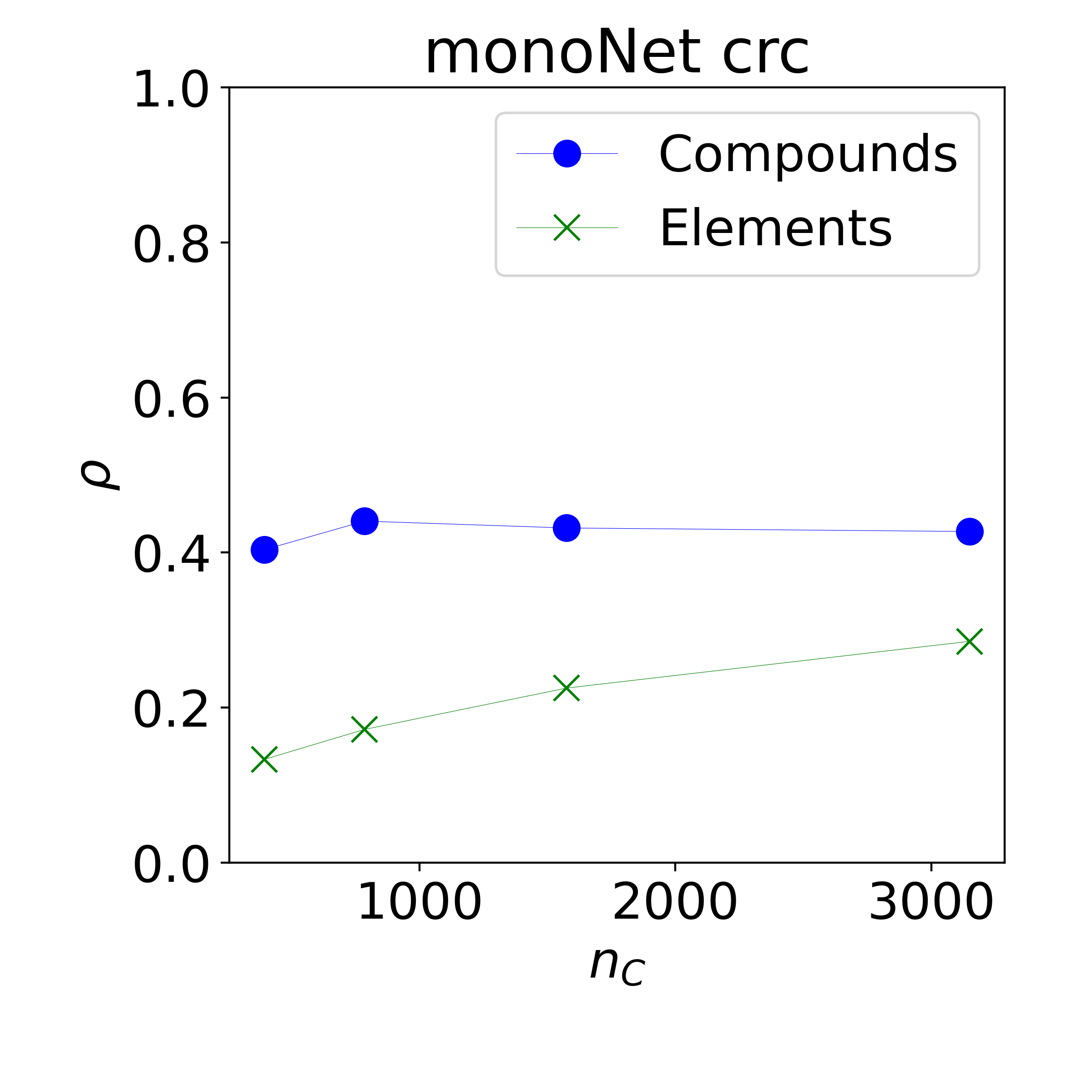}
\includegraphics[scale=0.5]{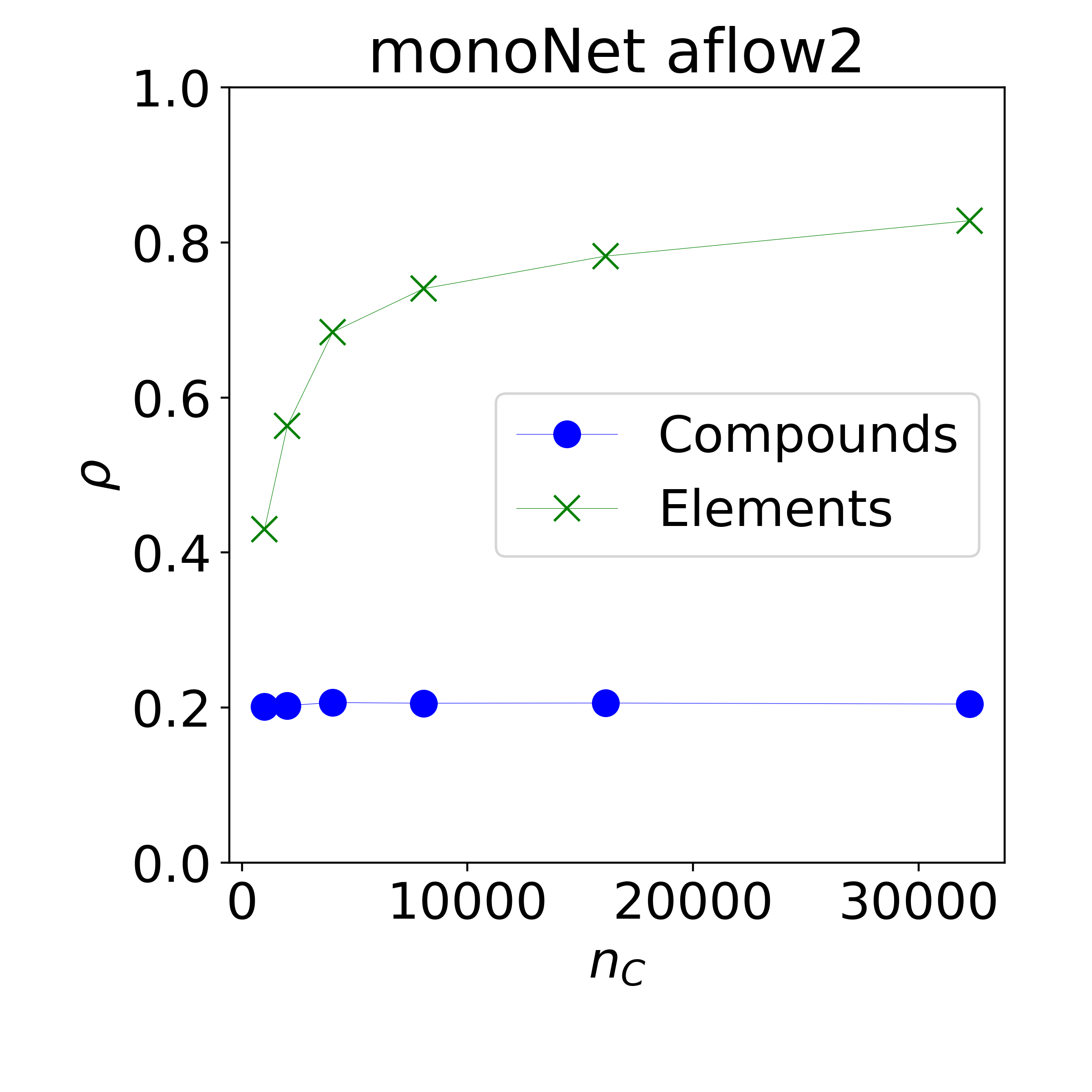}
\caption{Link density of the compounds, $\rho_C=L_C/(n_C(n_C-1)/2)$ (blue dots),  and elements, $\rho_E=L_E/(n_E(n_E-1)/2)$ (green x), networks, for the CRC (left) and AFLOW (right) datasets, plotted against dataset size of the number of compounds, $n_C$.
}
\label{fig:linkDensity}
\end{figure*}

The link density of the compounds networks saturates at a finite value, while the link density for the elements networks increases, as the number of compounds in the analyzed dataset increases, as shown in Figure \ref{fig:linkDensity}.

\subsection{Distributions of Normalized Degree}

\begin{figure*}[th]
\centering
\includegraphics[scale=0.5]{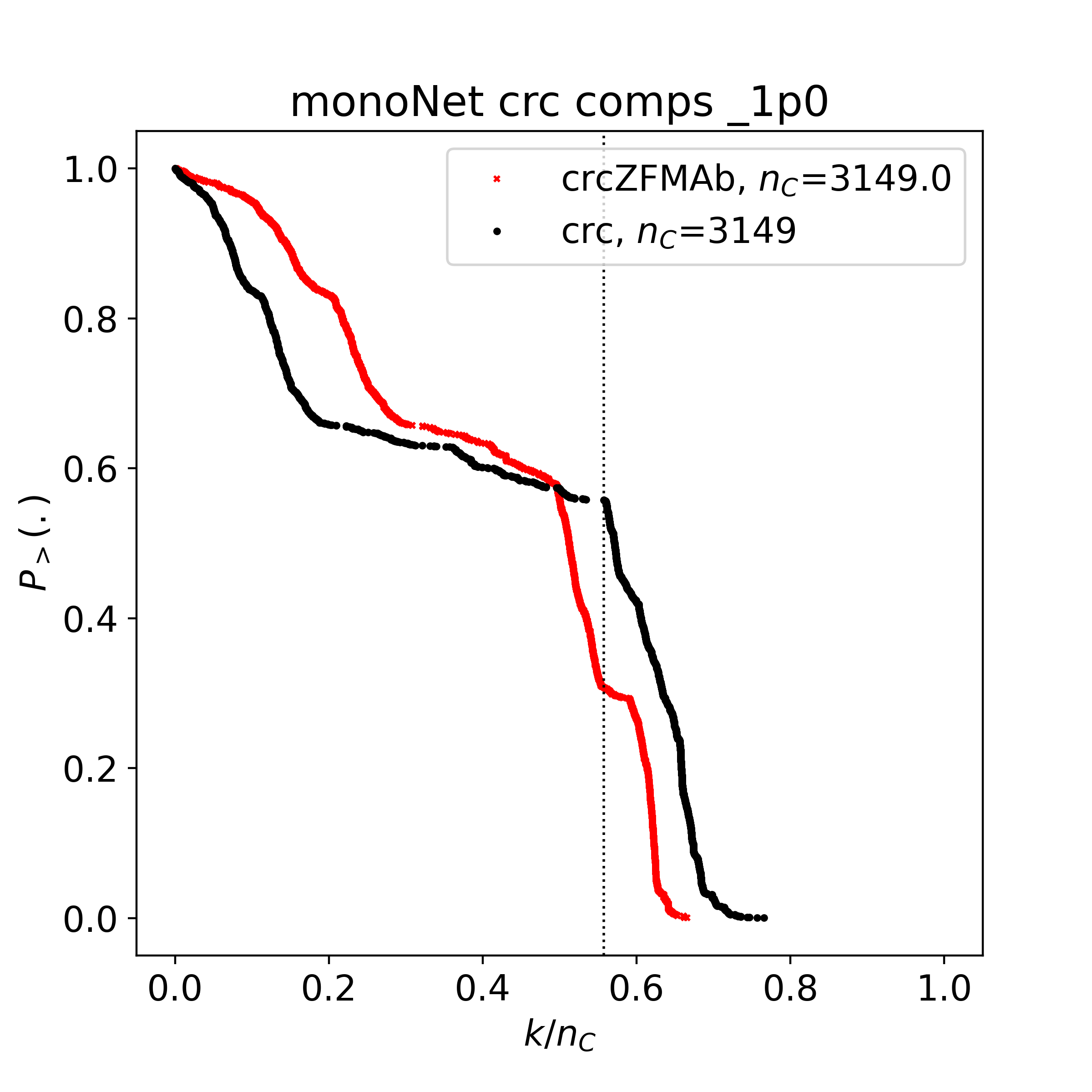}
\includegraphics[scale=0.5]{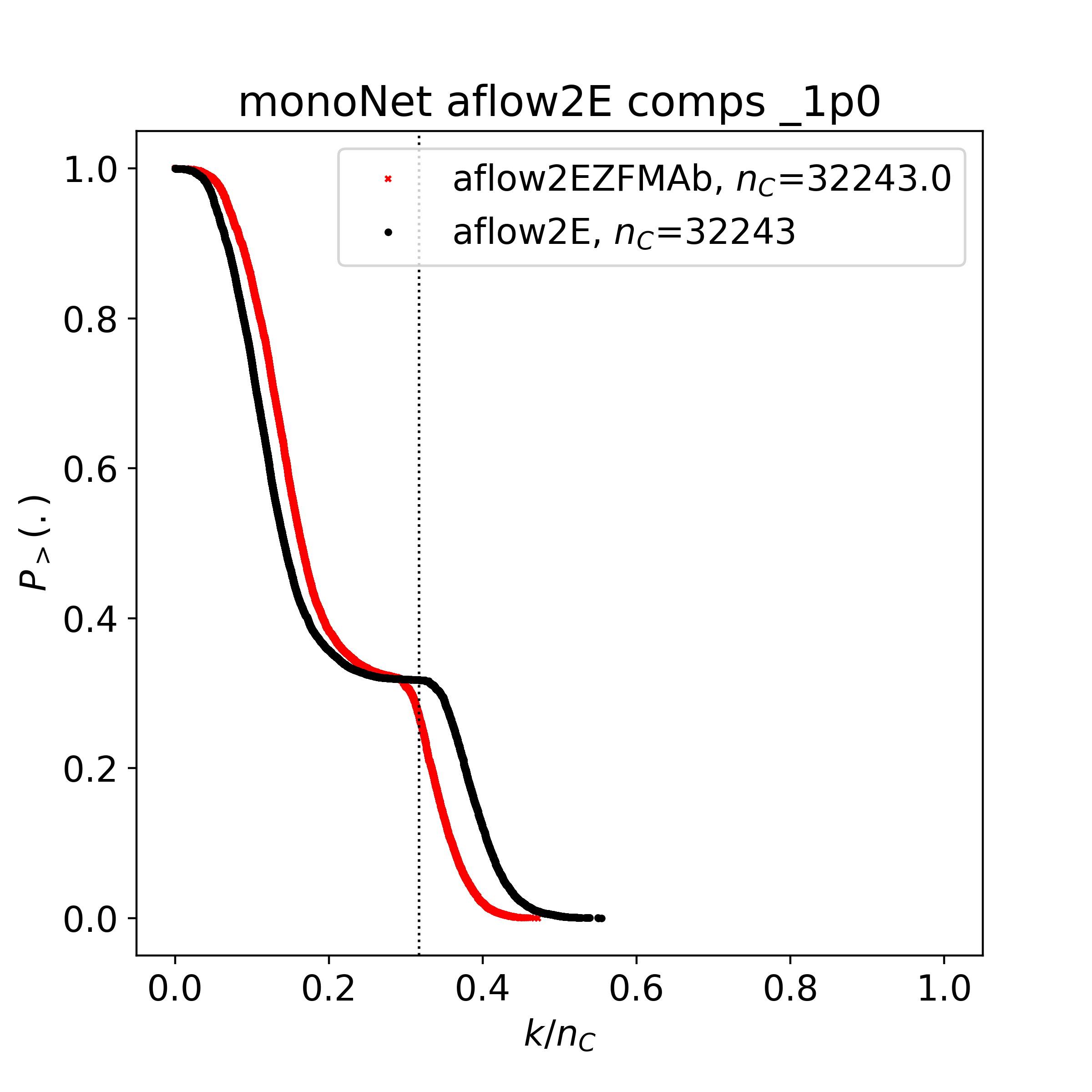}
\caption{Cumulative distributions of normalized degree, empirical (black dots) and calculated (red x), for the compounds of the CRC (left) and AFLOW (right) monopartite networks. As in Figure 2c,d of main, but the degrees are divided by the total number of nodes/compounds, $n_C$ in the dataset/network.
}
\label{fig:degNormDistm}
\end{figure*}

The distributions of normalized degrees by the total number of nodes decay faster for AFLOW than CRC. CRC has larger maximum degree relative to the maximum degree possible, as compared to AFLOW, as we show in Figure \ref{fig:degNormDistm}. 

\subsection{Degree assortativity of bipartite networks}

In the bipartite network the nearest neighbor degrees of the opposite layer are defined as
\begin{eqnarray}
d_c^{nn} = \frac{1}{d_c} \sum_{e=1}^{n_E} B_{ce} d_e, \qquad
d_e^{nn} = \frac{1}{d_e} \sum_{c=1}^{n_C} B_{ce} d_c
\end{eqnarray}

\begin{figure*}[th]
\centering
\includegraphics[scale=0.5]{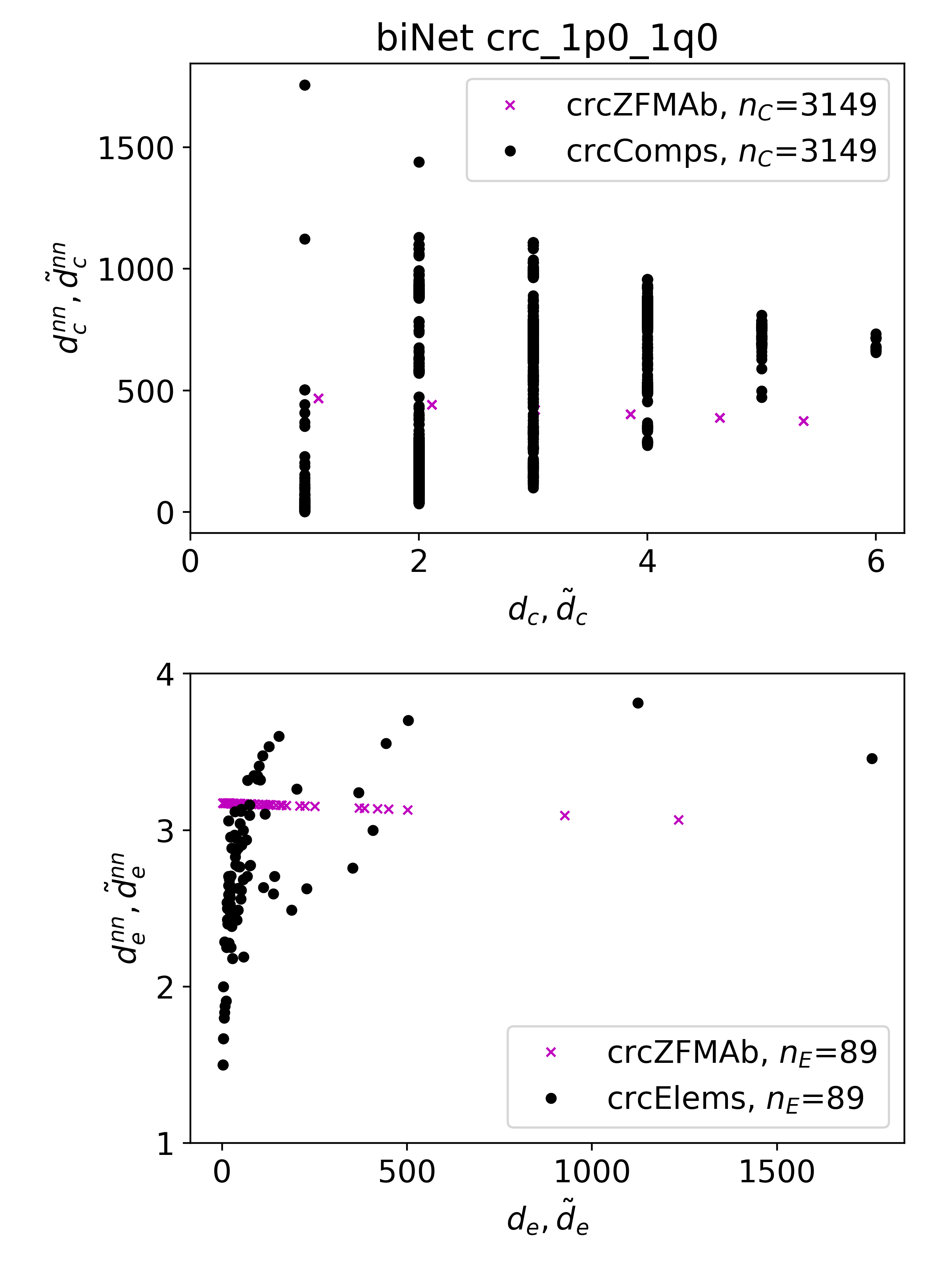}
\includegraphics[scale=0.5]{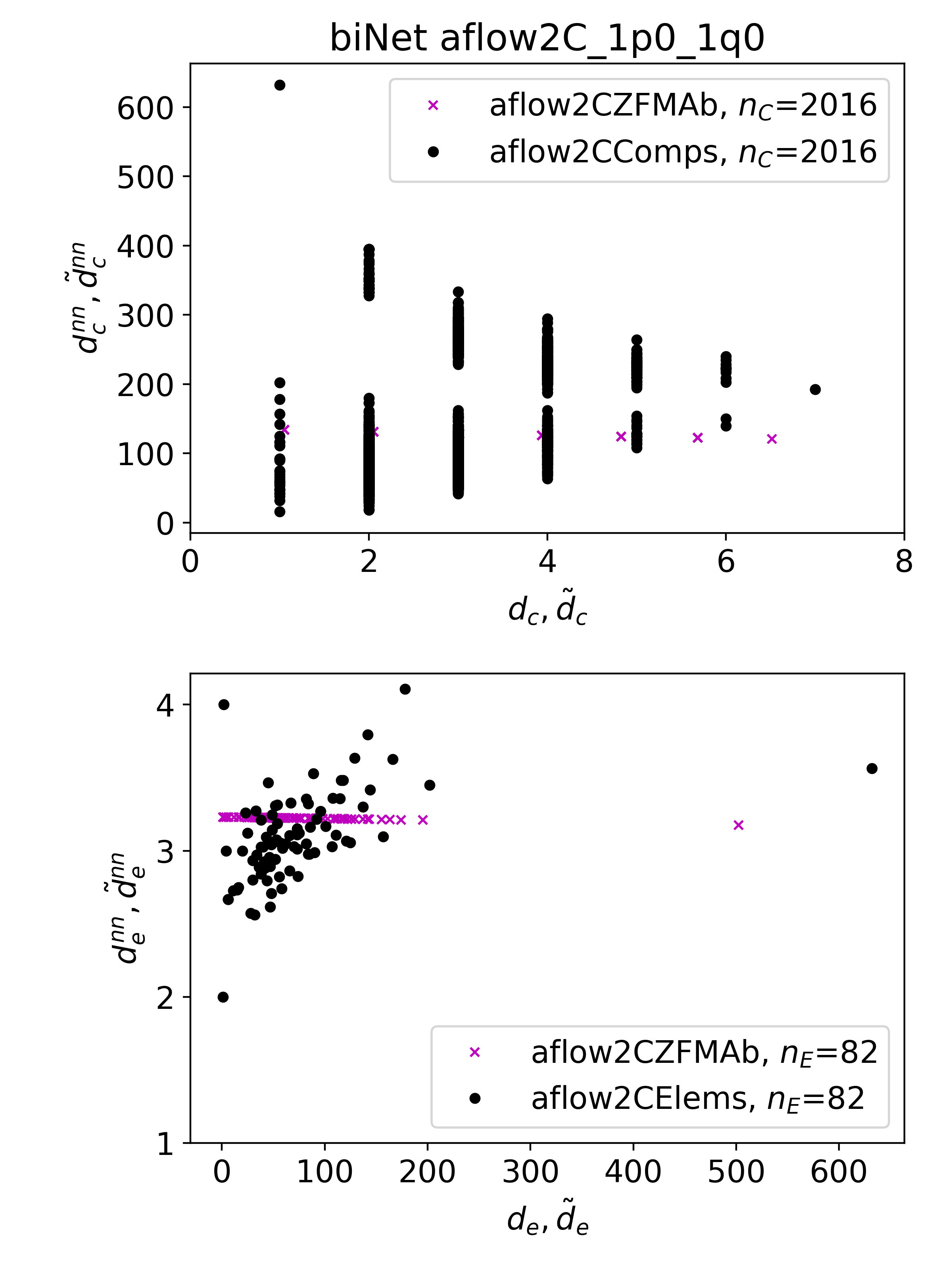}
\caption{Nearest neighbor degree vs degree, empirical (black dots) and calculated (purple x),  for the compounds (top) and elements (bottom) layers of the CRC (left) and AFLOW (right) bipartite networks. [An AFLOW dataset with a number of compounds of $n_C=2016$ was used for computational purposes.]
}
\label{fig:KnnvsK}
\end{figure*}

We can calculate the nearest neighbor degrees from the bipartite fitness model with a double sum
\begin{eqnarray}
\Tilde d_c^{nn} = \frac{1}{ \Tilde d_c} \sum_{e,c'} f(\delta^{*},x^b_e,y^b_c)f(\delta^*,x^b_e,y^b_{c'})\\
\Tilde d_e^{nn} = \frac{1}{ \Tilde d_e} \sum_{c,e'} f(\delta^{*},x^b_e,y^b_c)f(\delta^*,x^b_{e'},y^b_{c})
\end{eqnarray}
which are plotted alongside the empirical data in Fig.~\ref{fig:KnnvsK}. 

The degrees of the elements that are nearest neighbors to compounds, $d_{c}^{nn}$, are significantly spread so much so that they embrace both assortative and dis-assortative behaviors for the whole range of compounds degree, $d_{c}$, for both datasets, upper panels in Fig.~\ref{fig:KnnvsK}. The degrees of the compounds that are nearest neighbors to elements, $d_{e}^{nn}$, vs the degrees of the elements, $d_{e}$, appear as much more disordered clouds with a strong initial assortative behavior that dissipates or tends towards weak dis-assortativty for larger $d_{e}$ values, lower panels in Fig.~\ref{fig:KnnvsK}. Overall assortativity is captured in on average for both the elements and compounds layers in a partial success of our model of non-interacting fermions. 

\subsection{Degree assortativity of monopartite networks}

The nearest neighbor degree is defined as,
\begin{eqnarray}
k_c^{nn} = \frac{1}{ k_c } \sum_{c'} (A_C)_{cc'} k_{c'}, \qquad
k_e^{nn} = \frac{1}{ k_e } \sum_{e'} (A_E)_{ee'} k_{e'}
\end{eqnarray}

In the mono-partite network the nearest neighbor degrees are calculated as,
\begin{eqnarray}
\Tilde k_e^{nn} = \frac{1}{\Tilde k_e} \sum_{e',e''\neq e} f(\delta_E^{*},x^{m}_e,x^{m}_{e'}) f(\delta_E^{*},x^{m}_{e'},x^{m}_{e''}), \\
\Tilde k_c^{nn} = \frac{1}{\Tilde k_c} \sum_{c', c''\neq c)} f(\delta_C^{*},y^{m}_c,y^{m}_{c'})
f(\delta_C^{*},y^{m}_{c'},y^{m}_{c''})
\end{eqnarray}

\begin{figure*}[th]
\centering
\includegraphics[scale=0.5]{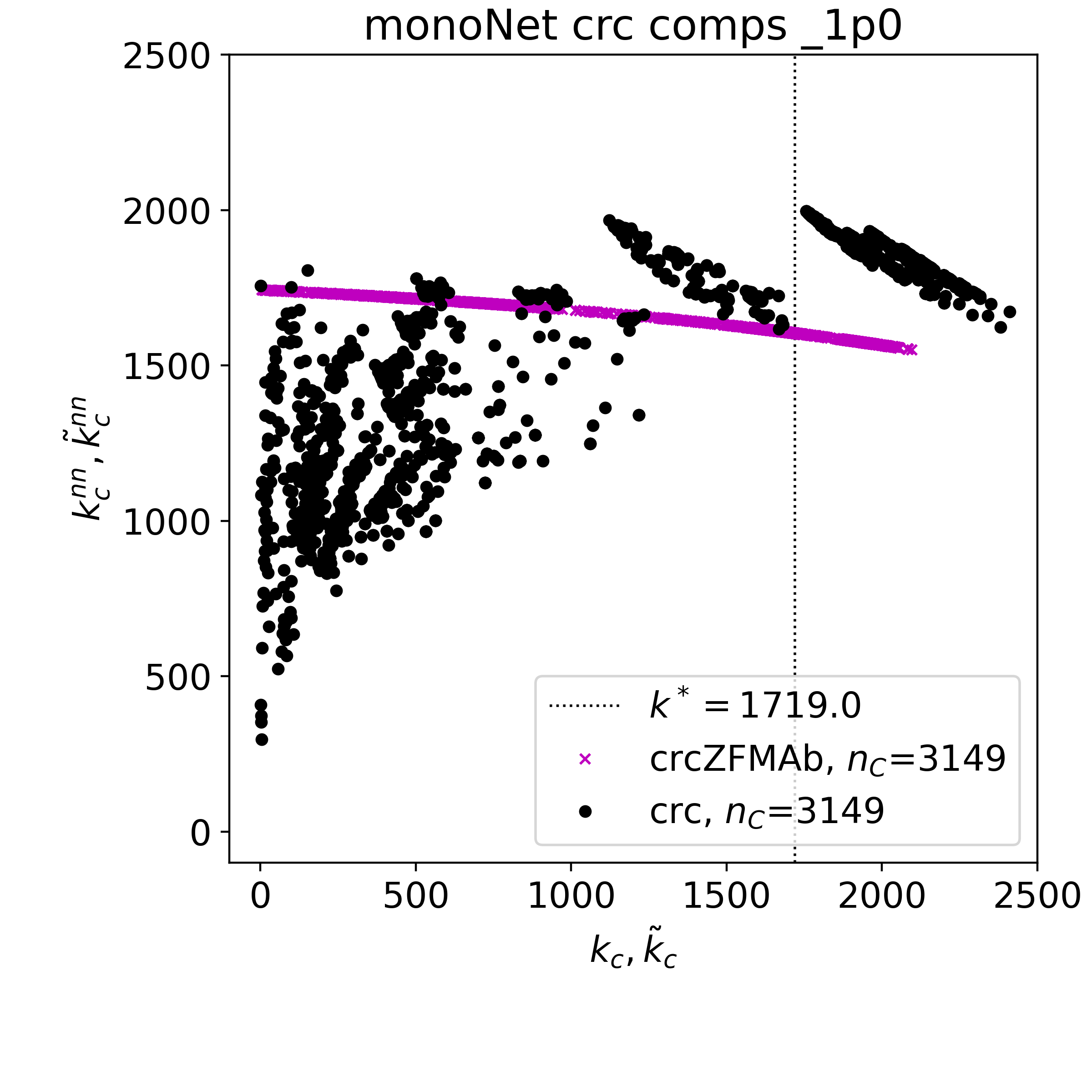}
\includegraphics[scale=0.5]{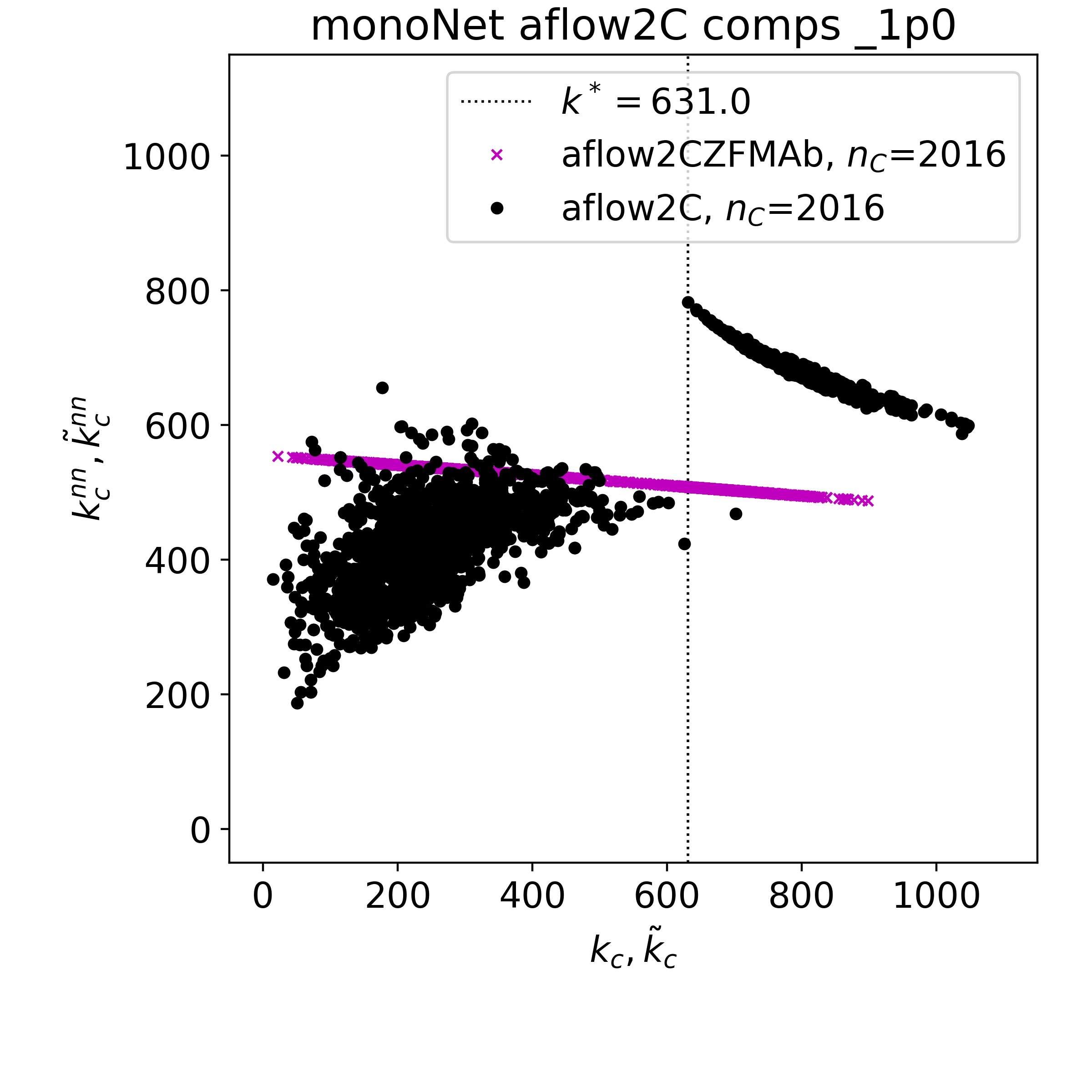}
\includegraphics[scale=0.5]{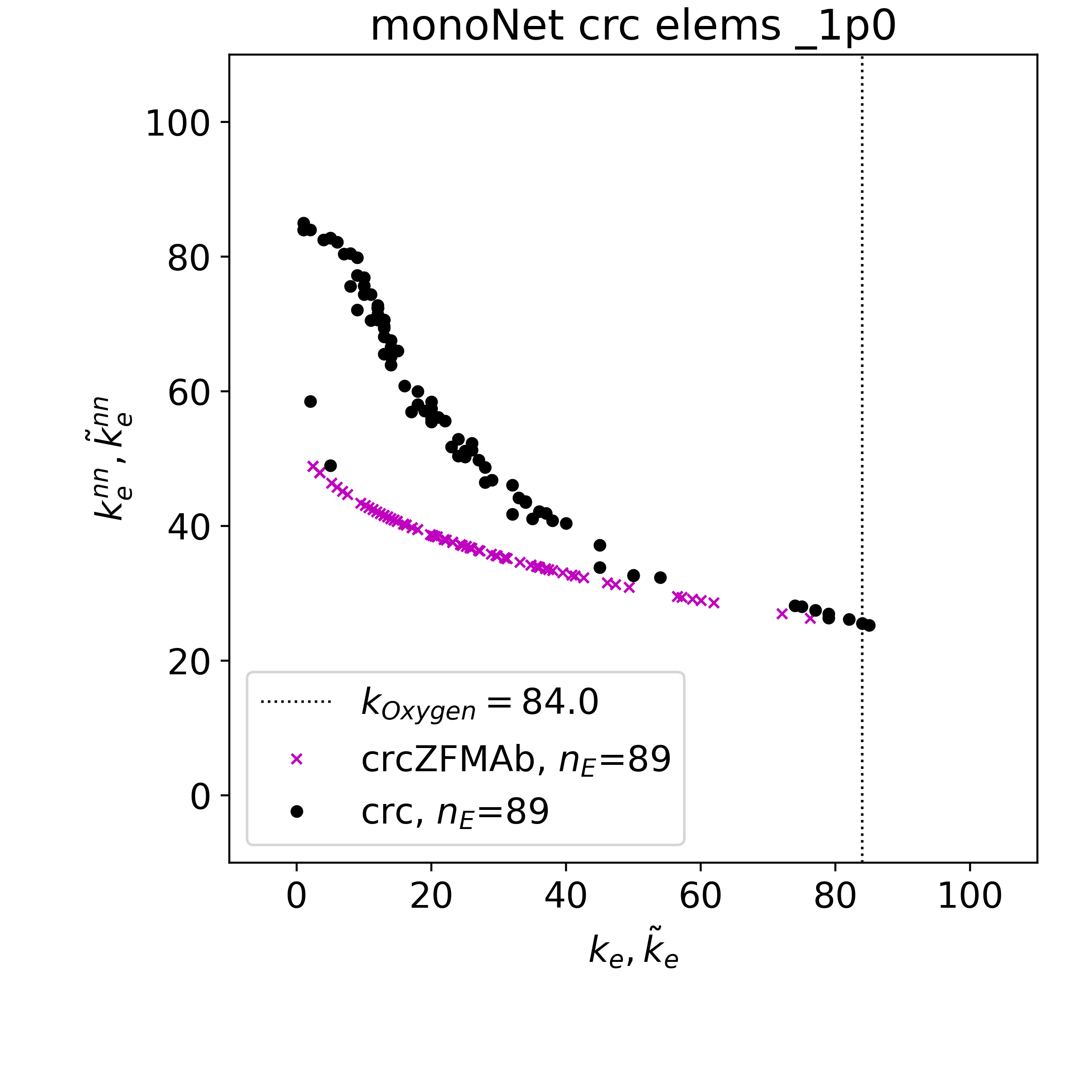}
\includegraphics[scale=0.5]{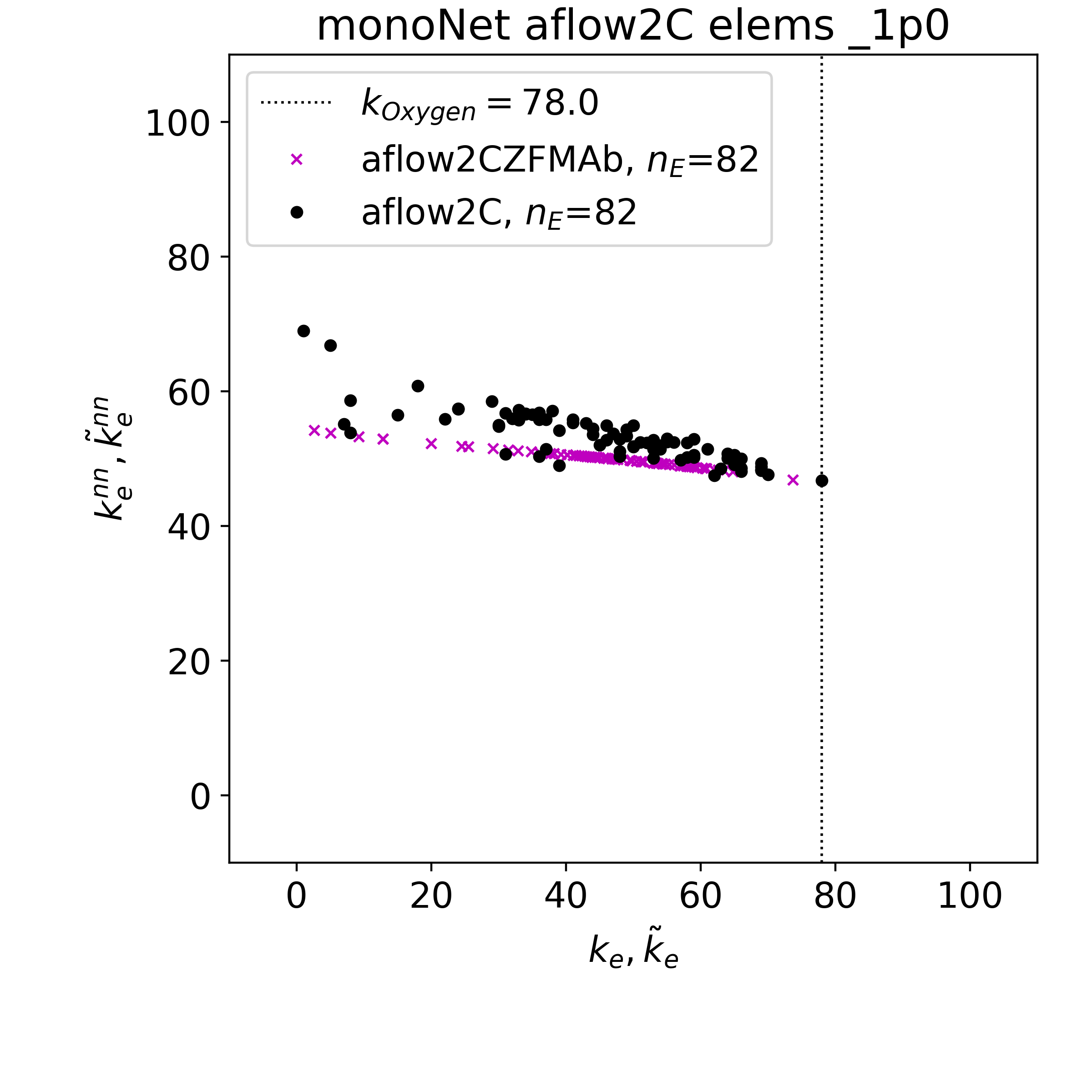}
\caption{Nearest neighbor degree vs degree, empirical (black dots) and calculated (purple x), for the compounds (top) and elements (bottom) of the mono-partite networks of CRC (left) and AFLOW  (right). [An AFLOW dataset with a number of compounds of $n_C=2016$ was used for computational purposes.]
}
\label{fig:KnnvsKm}
\end{figure*}

The nearest neighbors degrees are plotted alongside the empirical data in Figs. \ref{fig:KnnvsKm}. We find an agreement between the real networks and the fermionic fitness model as regards the dis-assortative behavior of the network of elements, Fig. \ref{fig:KnnvsKm}. The degree assortativity of the compounds networks is more complicated, with a cloud for smaller degrees, and a more linear behavior at larger degrees. The cloud  exhibits stronger assortative, rather than dis-assortative behavior, whereas the linear part is clearly dis-assortative. Our model has qualitatively more in common with the latter, and only captures the upper boundary of the former, in a partial success of our model of non-interacting fermions in the compounds networks.

\clearpage

\subsection{Modeled vs empirical degree of bipartite networks}

\begin{figure*}[h]
\centering
\includegraphics[scale=0.5]{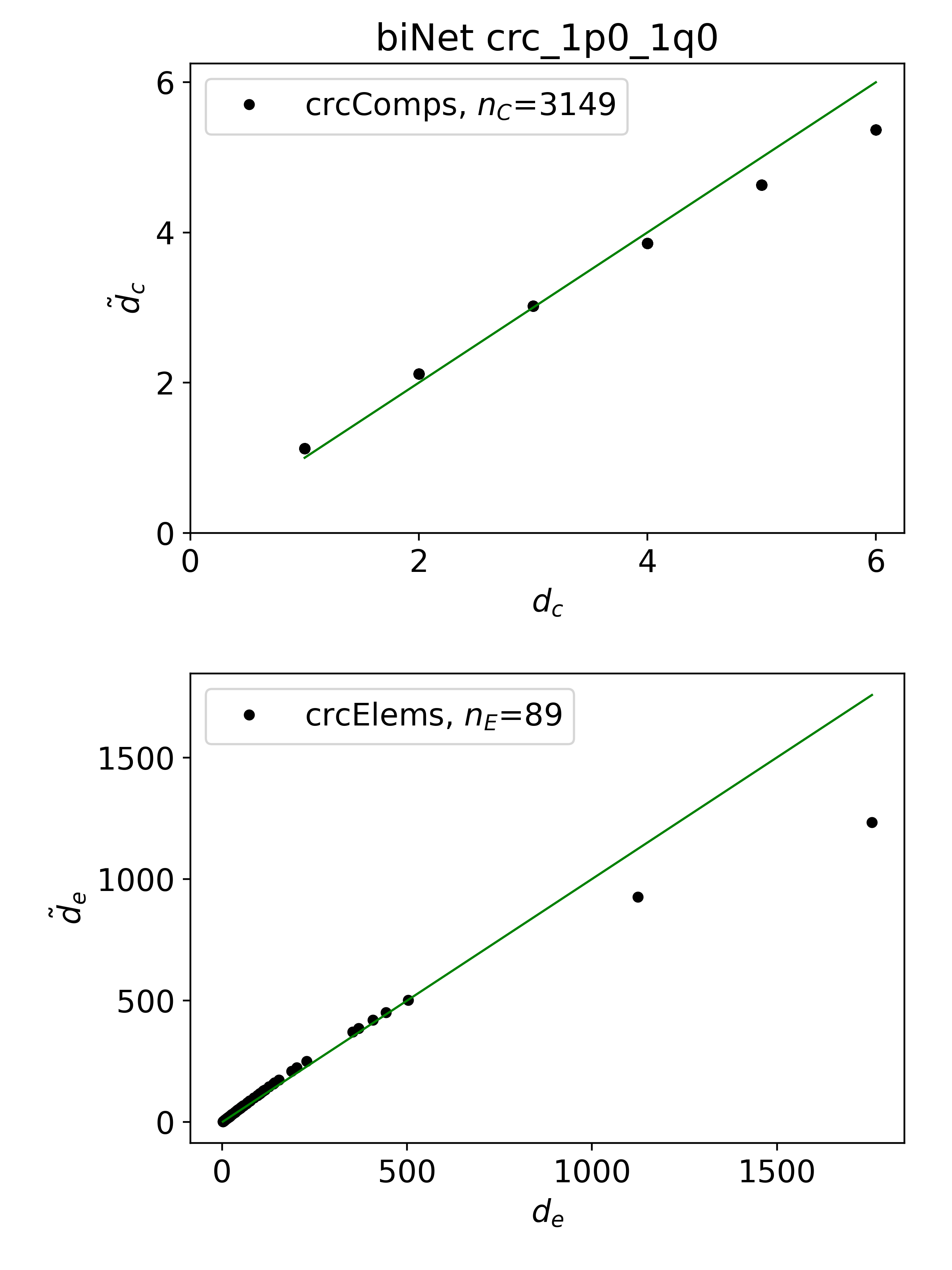}
\includegraphics[scale=0.5]{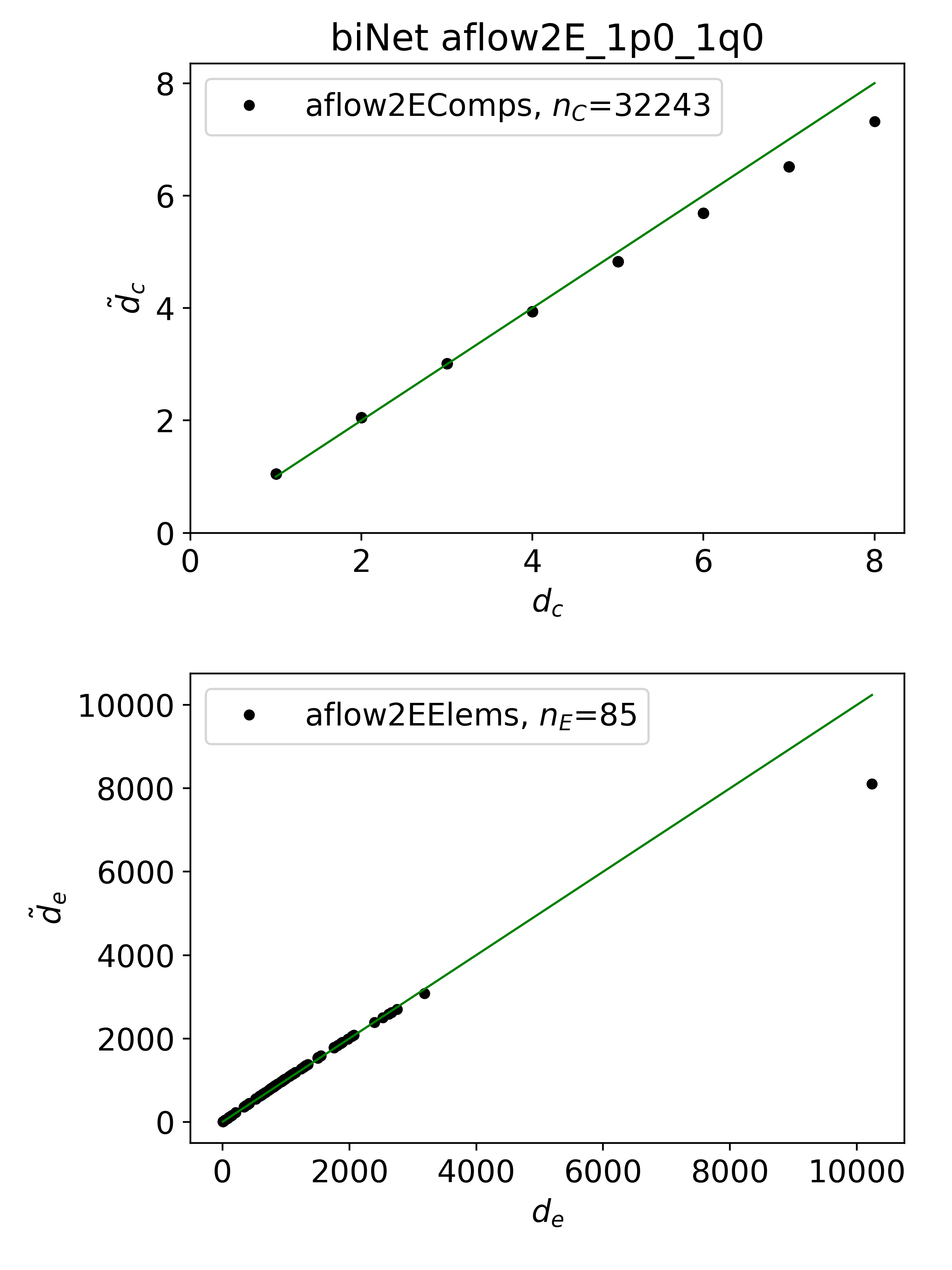}
\caption{The empirical degree vs the calculated degree (black dots), for the compounds (top) and elements (bottom) layers of the CRC (left), and AFLOW (right) bipartite networks. The green lines are diagonals $y=x$ as visual guides. 
}
\label{fig:KIKb}
\end{figure*}

\begin{figure*}[h]
\centering
\includegraphics[scale=0.5]{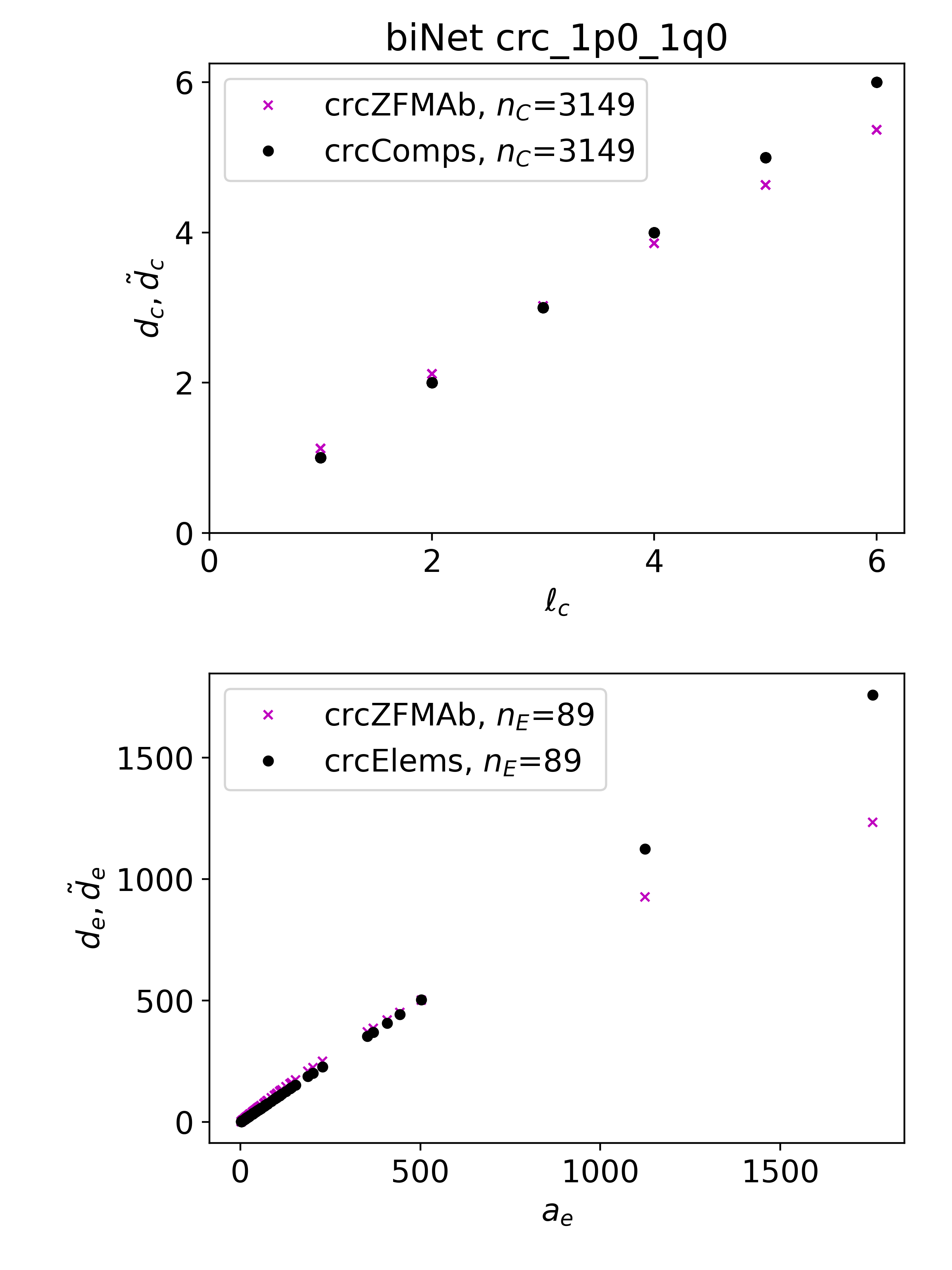}
\includegraphics[scale=0.5]{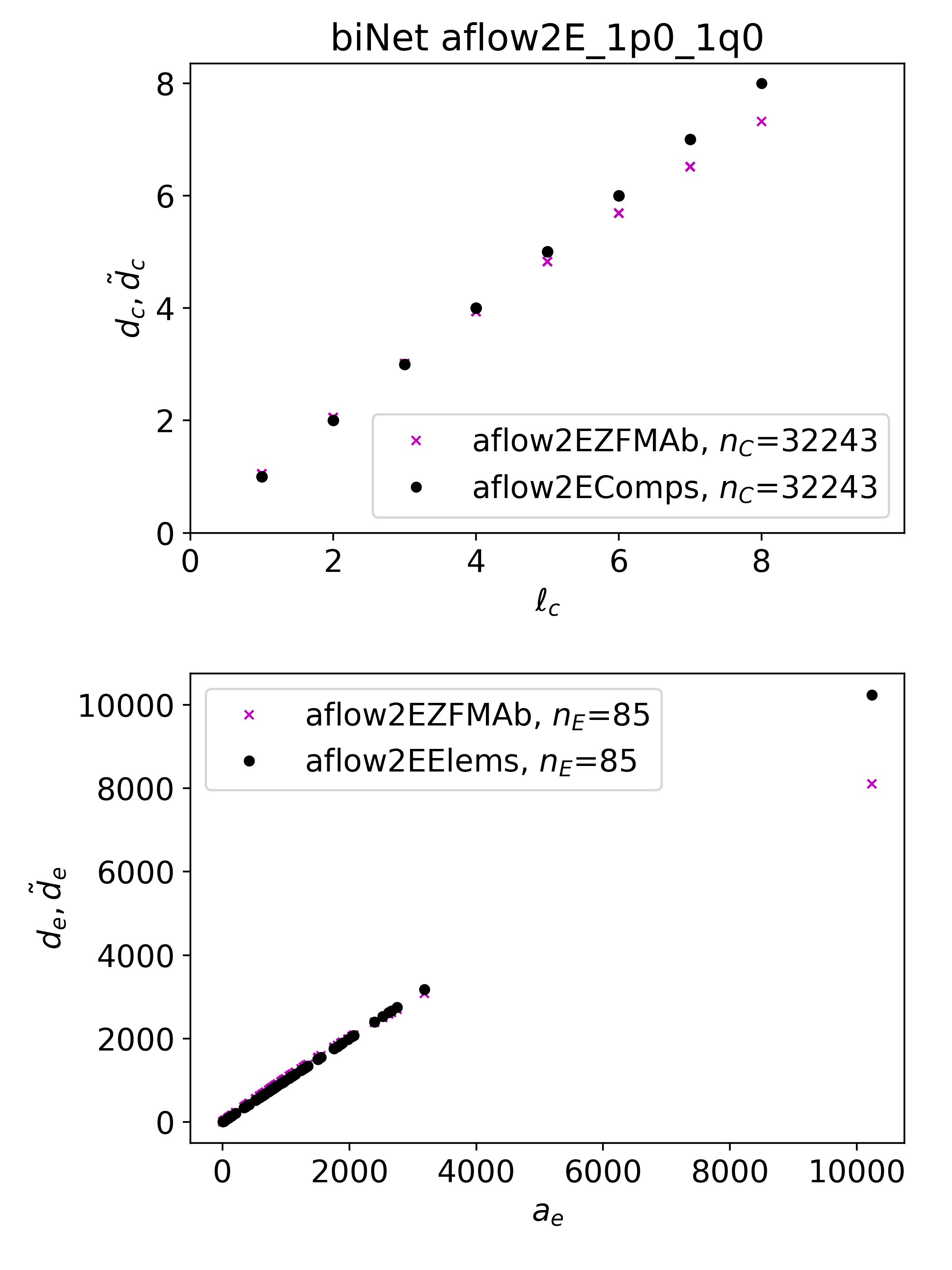}
\caption{The empirical degree (black dots) and calculated degree (purple x) vs fitness used in fermionic fitness model for the compounds (top) and elements (bottom) layers of the CRC (left), and AFLOW (right) bipartite networks.
}
\label{fig:KvFitb}
\end{figure*}

Below we show more figures with complementary results from the bipartite fitness modeling of the empirical networks. In Figure \ref{fig:KIKb} we show the calculated degrees versus the empirical degrees for both layers of the bipartite networks. In Figure \ref{fig:KvFitb} we show the empirical and modeled degrees versus the fitness for both layers of the bipartite networks.

\clearpage

\subsection{Degrees vs fitness of bipartite networks}

\clearpage

\subsection{Modeled vs empirical degree of monopartite networks}

\begin{figure*}[h]
\centering
\includegraphics[scale=0.5]{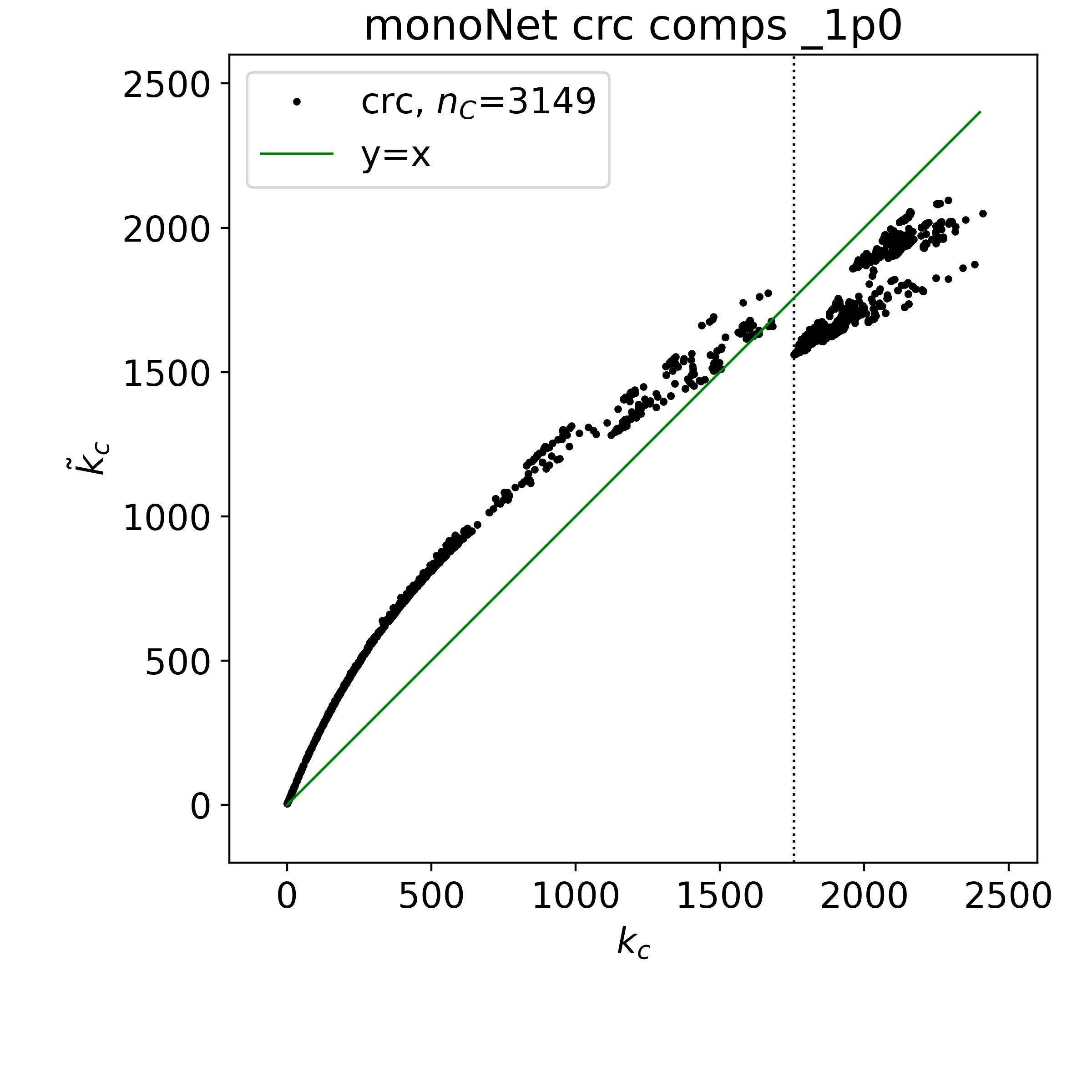}
\includegraphics[scale=0.5]{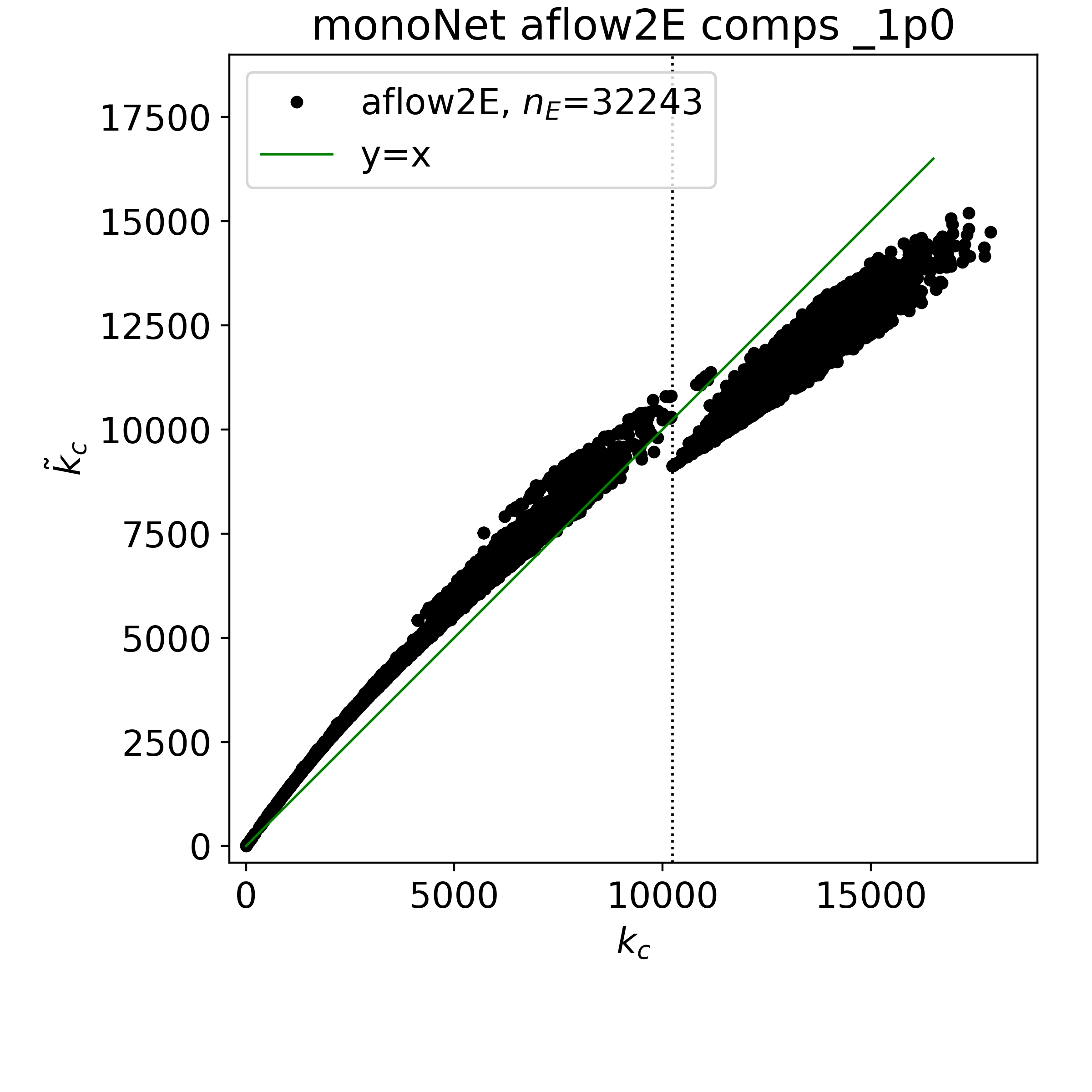}
\includegraphics[scale=0.5]{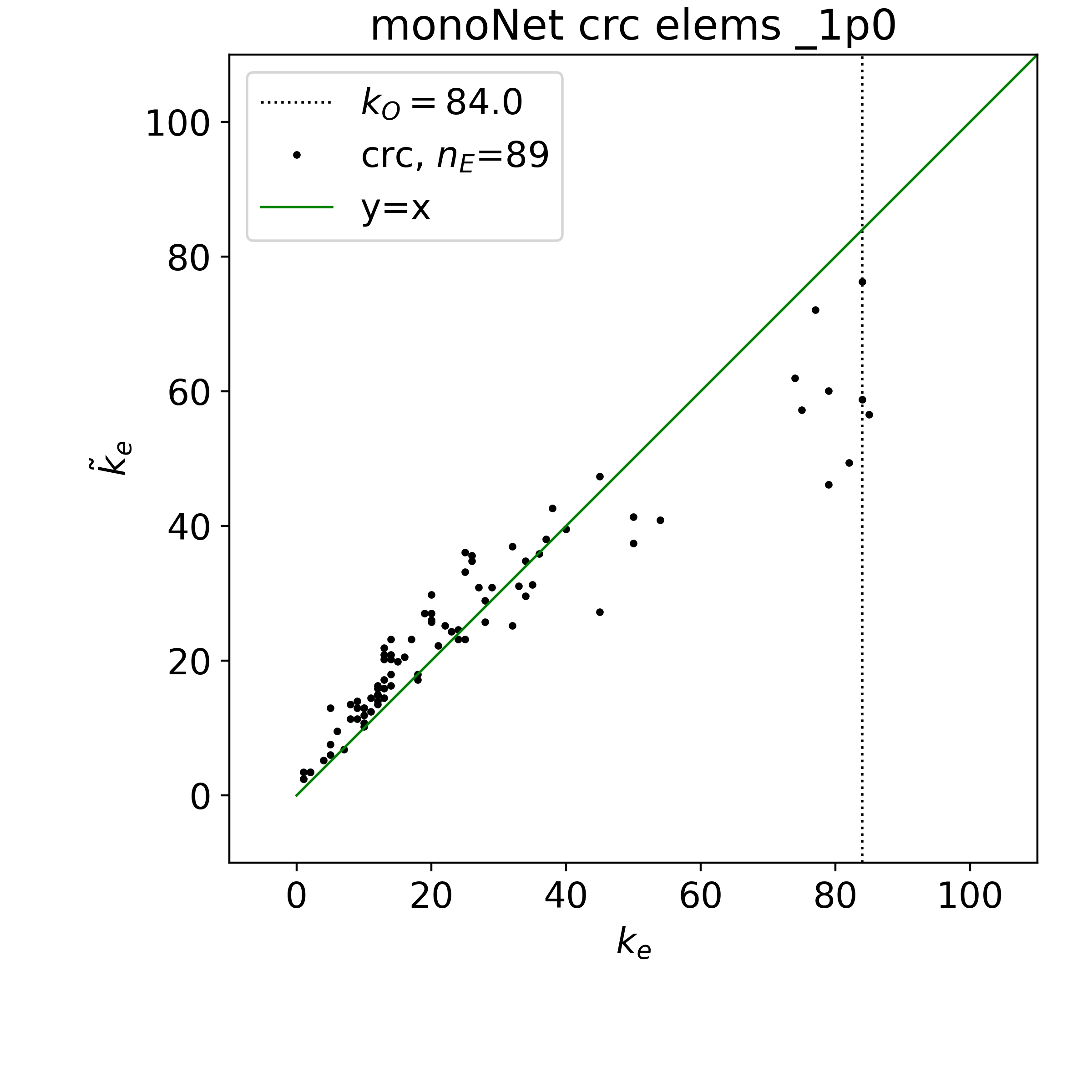}
\includegraphics[scale=0.5]{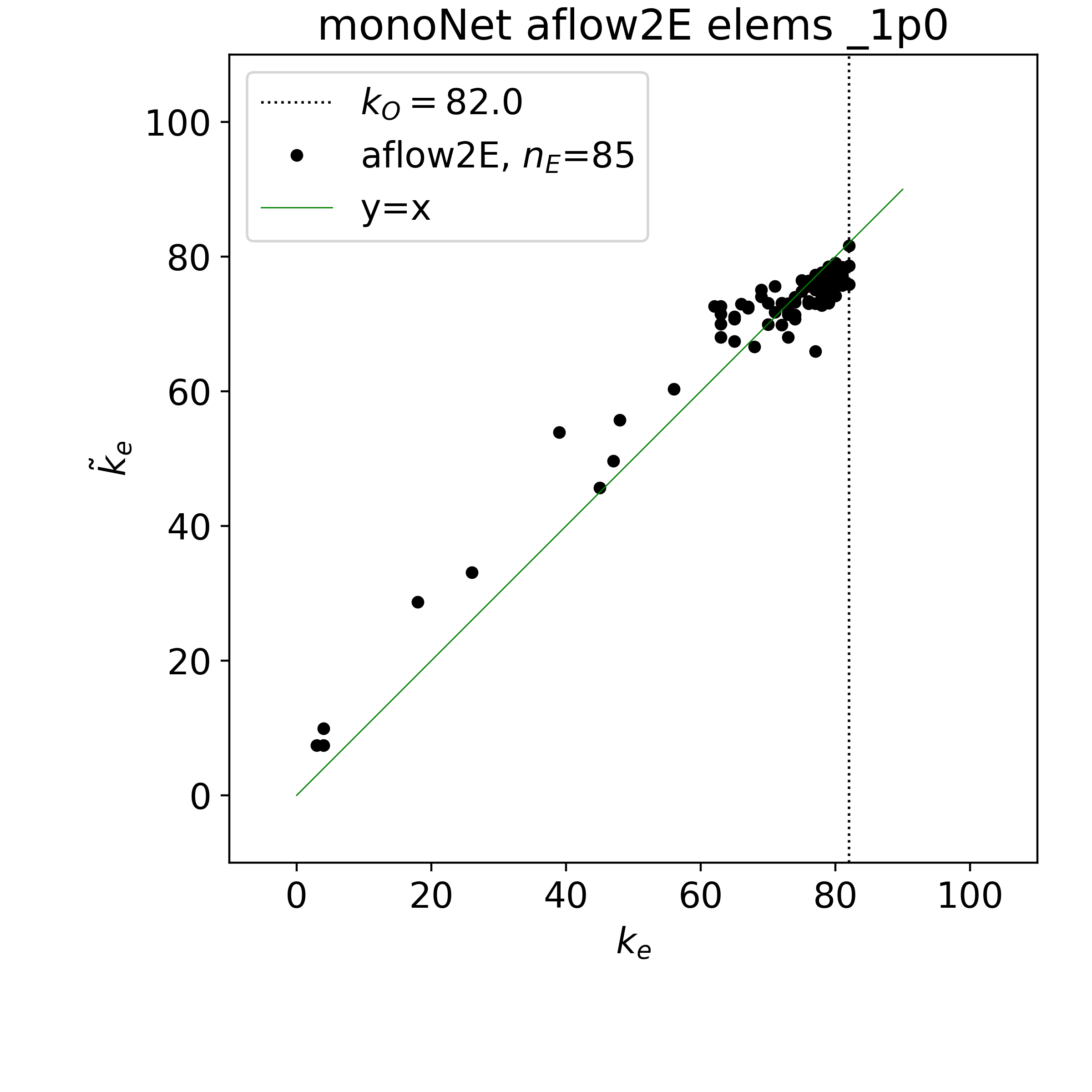}
\caption{The empirical degree vs the simulated degree (black dots) for the compounds (top) and elements (bottom) mono-partite networks of CRC (left) and AFLOW (right). The green lines are diagonals $y=x$ as visual guides.
}
\label{fig:KIKm}
\end{figure*}

\begin{figure*}[h]
\centering
\includegraphics[scale=0.5]{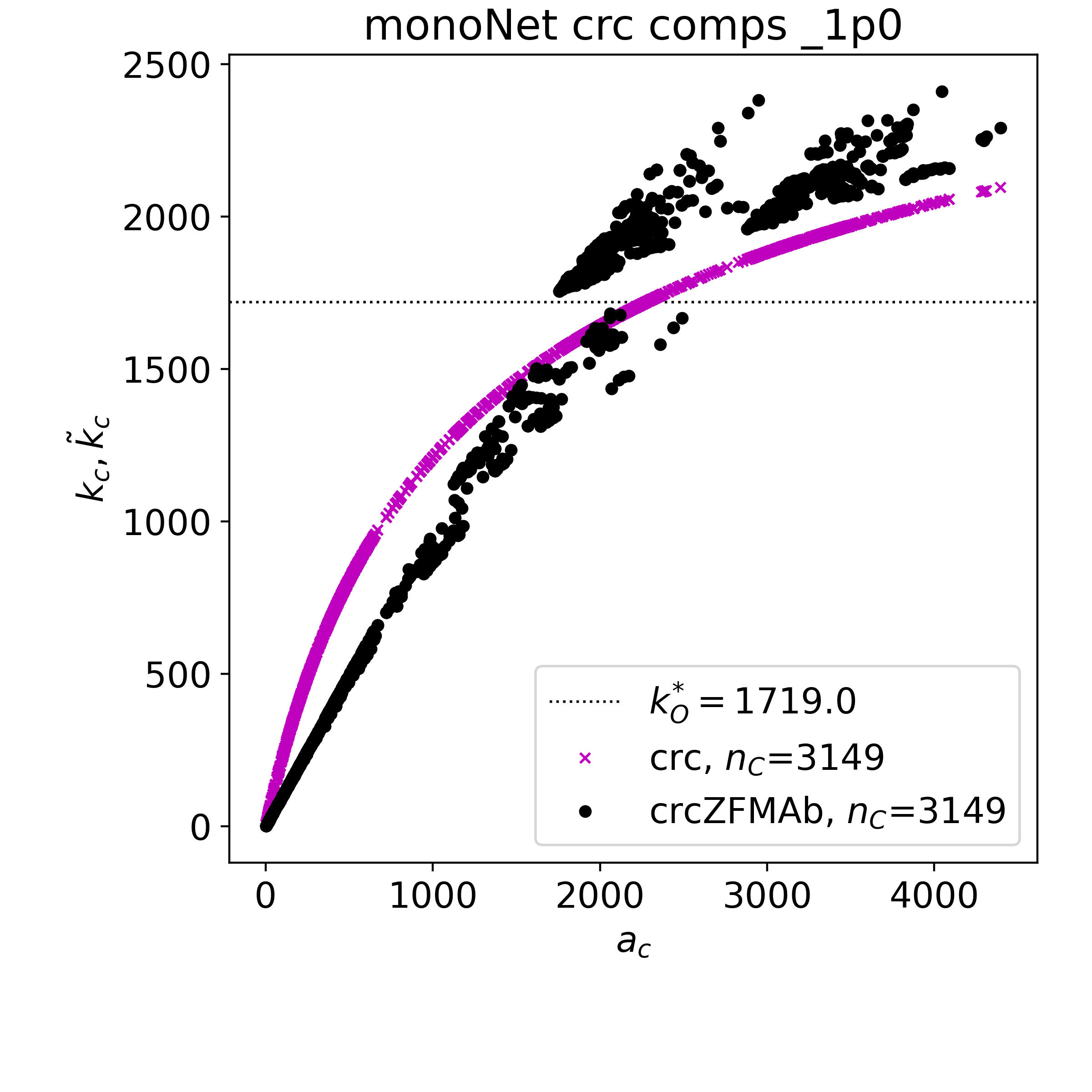}
\includegraphics[scale=0.5]{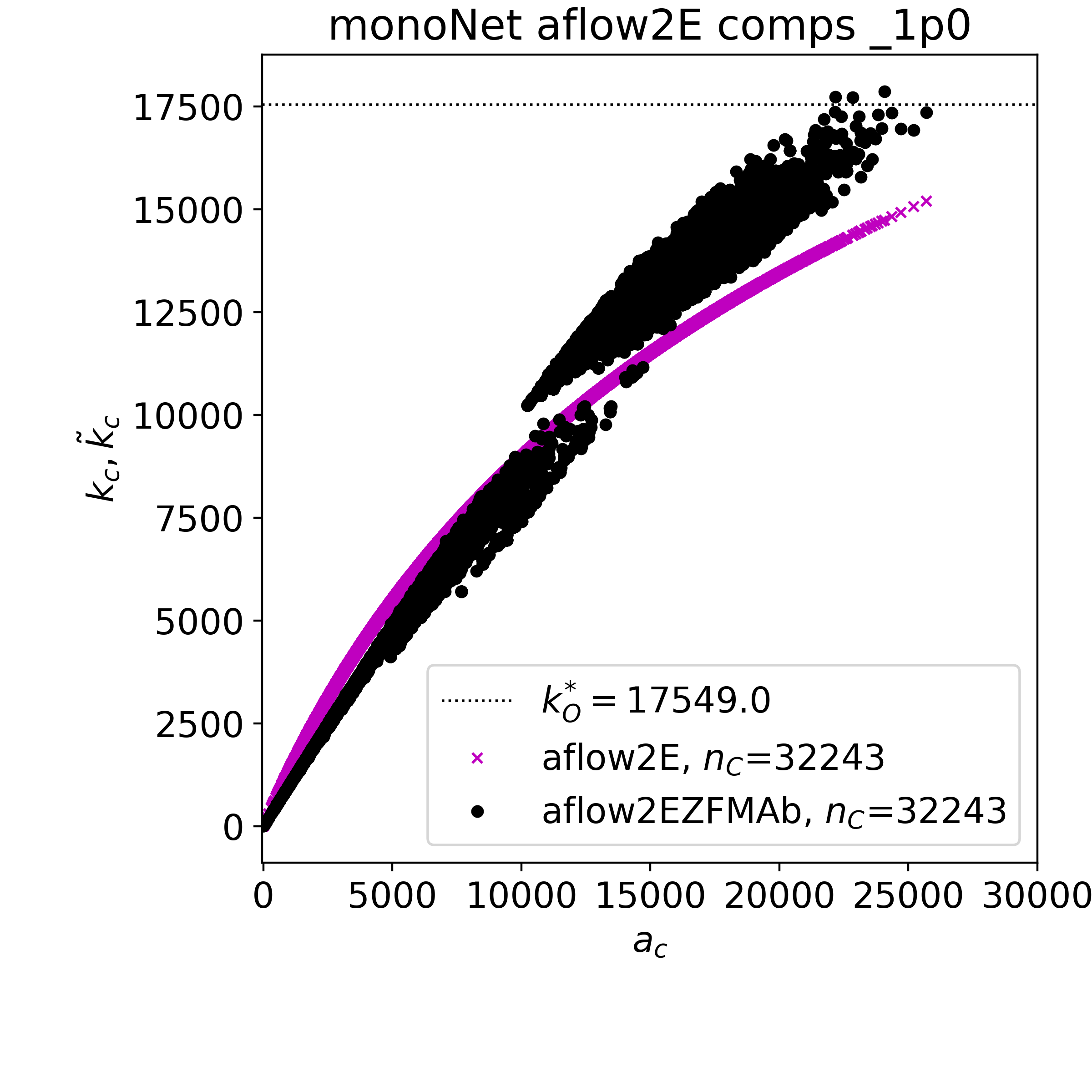}
\includegraphics[scale=0.5]{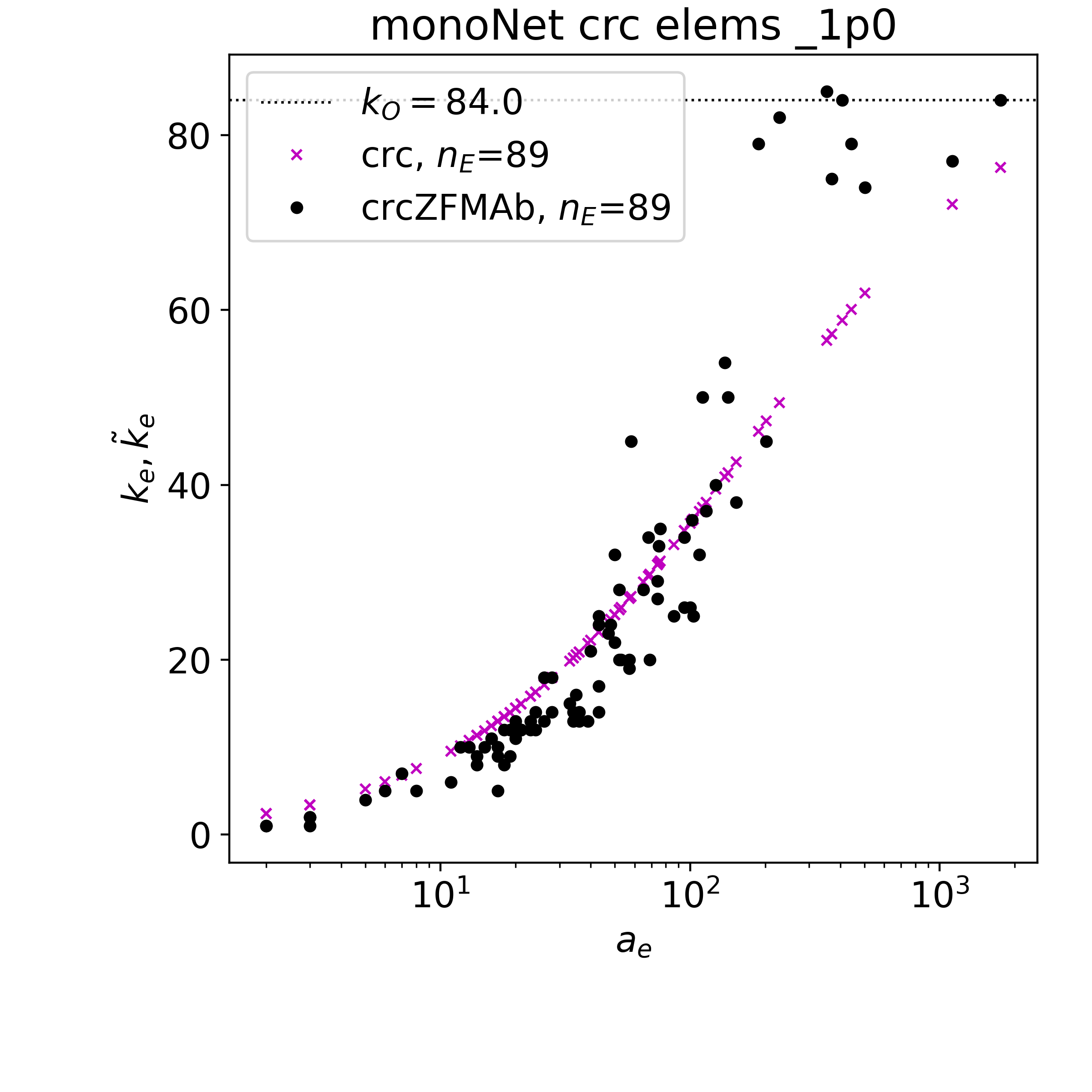}
\includegraphics[scale=0.5]{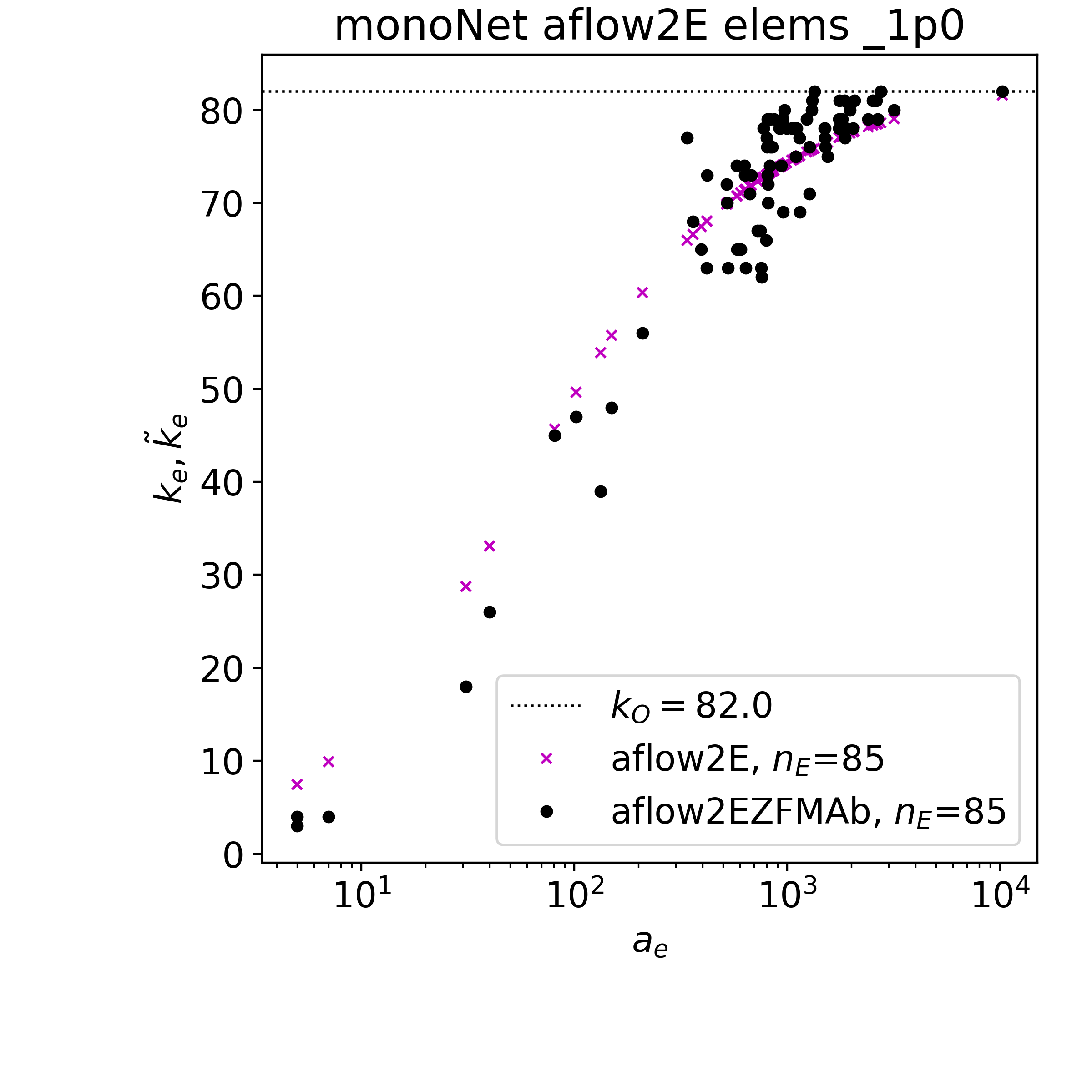}
\caption{The empirical degree (black dots) and simulated degree (purple x) vs fitness for the compounds (top) and elements (bottom) mono-partite networks of CRC (left) and AFLOW (right).
}
\label{fig:KvFitm}
\end{figure*}

Below we show more figures with complementary results from the monopartite fitness modeling of the empirical networks. In Figure \ref{fig:KIKm} we show the calculated degrees versus the empirical degrees for both monopartite networks. In Figure \ref{fig:KvFitm} we show the empirical and modeled degrees versus the fitness for both monopartite networks.

\clearpage

\subsection{Degrees vs fitness of monopartite networks}

\end{document}